\documentclass[epj,final]{svjour}
\pdfoutput=1
\usepackage{latexsym}
\usepackage{graphicx}
\usepackage[dvipsnames]{xcolor}
\usepackage{xspace}
\usepackage{amsmath}
\usepackage{amsfonts}
\usepackage{amssymb}
\usepackage{bm}
\usepackage[colorlinks=true,linkcolor=blue,citecolor=blue,pdfauthor=Gross]{hyperref}

\begin{document}
\title{Josephson Coupling and Fiske Dynamics in Ferromagnetic Tunnel Junctions}
\titlerunning{Josephson Coupling and Fiske Dynamics in Ferromagnetic Tunnel Junctions}
 \author{G.~Wild\inst{1,2}
    \and C.~Probst\inst{1}
    \and A.~Marx\inst{1}
    \and R. Gross\inst{1,2}
}
\authorrunning{G. Wild et al.}
\institute{Walther-Mei{\ss}ner-Institut, Bayerische Akademie der Wissenschaften,
           Walther-Mei{\ss}ner-Stra{\ss}e 8, D-85748 Garching, Germany.
      \and Physik-Department, Technische Universit\"{a}t M\"{u}nchen,
           James-Franck-Stra{\ss}e, D-85748 Garching, Germany.
 }
\date{Received: date} %  / Revised version: date}
\mail{Rudolf.Gross@wmi.badw.de}

\abstract{ We report on the fabrication of Nb/AlO$_x$/$\rm
Pd_{0.82}Ni_{0.18}$/Nb superconductor/insulator/ferromagnetic
metal/superconductor (SIFS) Josephson junctions with high critical current
densities, large normal resistance times area products, high quality factors,
and very good spatial uniformity. For these junctions a transition from 0- to
$\pi$-coupling is observed for a thickness $d_{\rm F} \simeq 6$\,nm of the
ferromagnetic $\rm Pd_{0.82}Ni_{0.18}$ interlayer. The magnetic field
dependence of the $\pi$-coupled junctions demonstrates good spatial homogeneity
of the tunneling barrier and ferromagnetic interlayer. Magnetic
characterization shows that the $\rm Pd_{0.82}Ni_{0.18}$ has an out-of-plane
anisotropy and large saturation magnetization, indicating negligible dead
layers at the interfaces. A careful analysis of Fiske modes provides
information on the junction quality factor and the relevant damping mechanisms
up to about 400\,GHz. Whereas losses due to quasiparticle tunneling dominate at
low frequencies, the damping is dominated by the finite surface resistance of
the junction electrodes at high frequencies. High quality factors of up to 30
around 200\,GHz have been achieved. Our analysis shows that the fabricated
junctions are promising for applications in superconducting quantum circuits or
quantum tunneling experiments.
}

\maketitle
\sloppy

\section{Introduction}
 \label{sec:Introduction}

 The interplay between superconductivity (S) and ferromagnetism (F) at
 S/F-interfaces and more complex S/F-heterostructures results in a rich variety
 of interesting
 phenomena~\cite{FuldePRA1964,Larkin1964,Bulaevskii1977,BuzdinRMP2004,Buzdin1982,Buzdin1992}.
 Examples are the oscillations in the critical temperature of S/F-bilayers with
 increasing F thickness~\cite{Buzdin1982,JiangPRL1996}, oscillations in the
 critical current of SFS Josephson
 junctions~\cite{RyazanovPRL2001,KontosPRL2002,SellierPRB2003,RyazanovJLTP2004,ObonzovPRL2006},
 or the variation of the critical temperature of FSF trilayers as a function of
 the relative magnetization direction in the F layers~\cite{GuPRL2002}. At an
 S/F-interface, the superconducting order parameter does not only decay in the F
 layer as at superconductor/normal metal (S/N) interfaces but shows a spatial
 oscillation resulting in a sign change. This oscillatory behavior is the direct
 consequence of the exchange splitting of the spin-up and spin-down subbands in
 the F layer, causing a finite momentum shift $q=\pm 2E_{\rm ex}/v_{\rm F}$ of
 the spin-up and spin-down electron of a Cooper pair leaking into the F
 layer~\cite{BuzdinRMP2004,Buzdin1982,Buzdin1992}. Here, $E_{\rm ex}$ is the
 exchange energy and $v_{\rm F}$ the Fermi velocity in the F layer. The decay
 and oscillatory behavior of the order parameter in the F layer can be described
 by the complex coherence length $\xi_{\rm F}^{-1} = \xi_{\rm
   F1}^{-1}+\imath\xi_{\rm
   F2}^{-1}$~\cite{BuzdinRMP2004,Buzdin1982,Buzdin1992,DemlerPRB1997,BuzdinPRB2003,BergeretPRB2003,BergeretPRB2007,FaurePRB2006}.
 For the case of large exchange energy and negligible spin-flip scattering,
 $\xi_{\rm F1}=\xi_{\rm F2} =\sqrt{\hbar D/E_{\rm ex}}$ with $D$ the diffusion
 coefficient in the F layer and $\hbar=h/2\pi$ the reduced Planck
 constant~\cite{BuzdinRMP2004,Buzdin1992,BuzdinPRB2003}. The oscillatory
 behavior of the order parameter has been unambiguously proven by the
 observation of the $\pi$-coupled state in SFS Josephson
 junctions~\cite{RyazanovPRL2001,KontosPRL2002,SellierPRB2003,RyazanovJLTP2004,ObonzovPRL2006,Guichard2003,Blum2002,Bauer2004,Frolov2004,Robinson2006,WeidesAPL2008,PfeifferPRB2008,Petkovic2009,Khaire2009}.
 Here, the ground state has a phase difference of $\pi$ between the macroscopic
 superconducting wave functions in the junction electrodes. Therefore,
 $\pi$-junctions have an anomalous current-phase relation $I_{\rm s} = I_{\rm c}
 \sin (\varphi +\pi) = -I_{\rm c} \sin\varphi$ corresponding to a negative
 critical current $-I_c$~\cite{Bulaevskii1977,DemlerPRB1997}. Using Josephson
 junctions with a step-like F layer thickness, also junctions with a coupling
 changing between $0$ and $\pi$ along the junction have been
 realized~\cite{BuzdinPRB2003b,WeidesPRL2006,WeidesJAP2007}.

An interesting application of $\pi$-coupled Josephson junctions are $\pi$-phase
shifters in superconducting quantum circuits. For example, superconducting flux
quantum
bits~\cite{Orlando1999,Chiorescu2003,DeppePRB2007,Deppe2008a,Niemczyk2009a,Niemczyk2010a}
require an external flux bias $\Phi_{\rm ext}=\Phi_0(n+\frac{1}{2})$ to operate
them at the degeneracy point. Here, $\Phi_0=h/2e$ is the flux quantum with the
elementary charge $e$ and $n$ is an integer. This requirement makes flux qubits
susceptible to flux noise, which may be introduced through the flux biasing
circuitry. Furthermore, to operate a cluster of (coupled) flux qubits, which
inevitably will have a spread in parameters, requires an individual and precise
flux bias for each qubit. To circumvent this problem the insertion of
$\pi$-phase shifters into the flux qubit loop has been
suggested~\cite{BlatterPRB2001,Yamashita2005,Yamashita2006,Ioffe1999} and
successfully demonstrated recently~\cite{Feofanov2010}. The applications of
$\pi$-coupled Josephson junctions in classical or quantum circuits in most cases
requires high critical current densities $J_{\rm c}$ and high critical current
times normal resistance products, $I_{\rm c}R_{\rm n}$. Furthermore, they are
not allowed to deteriorate the coherence properties when used in superconducting
quantum circuits. Potential sources of decoherence are for instance spin-flip
processes in the F layer or the dynamic response of the magnetic domain
structure. Another serious difficulty comes from dissipation due to
quasiparticle excitation. It has been shown recently, that long decoherence
times require junctions with both high $J_{\rm c}$ and large normal resistance
times area products, $R_{\rm n}\cdot A$~\cite{KatoPRB2007}. This is difficult to
be achieved in SFS junctions due to the low resistivity of the metallic F
layer. However, the situation can be improved by inserting an additional
insulating barrier, resulting in a
superconductor/insulator/ferromagnet/superconductor (SIFS) stack. Here, much
higher $R_{\rm n}\cdot A$ values at modest $J_{\rm c}$ can be achieved. In
particular, underdamped SIFS Josephson junctions can be realized allowing for
the study of the junctions dynamics. Therefore, there has been strong interest
in SIFS junctions recently~\cite{PfeifferPRB2008,Petkovic2009}.

In this article we report on the fabrication as well as the static and dynamic
properties of a series of Nb/AlO$_x$/$\rm Pd_{0.82}Ni_{0.18}$/Nb (SIFS)
Josephson junctions with different thickness $d_{\rm F}$ of the ferromagnetic
interlayer. We succeeded in the fabrication of junctions with high $J_{\rm c}$
and $R_{\rm n}\cdot A$ values resulting in high junction quality factors $Q$.
We discuss the dependence of the $I_{\rm c}R_{\rm n}$ product on the thickness
of the F layer, including the transition from $0$- to $\pi$-coupling. We also
address the magnetic properties of the F interlayer in the magnetic field
dependence of the critical current. Special focus is put on the analysis of
Fiske modes. Comparing these resonant modes to theoretical models allows us to
evaluate the quality factors of the junctions over a wide frequency range from
about 10 to 400\,GHz and to determine the dominating damping mechanisms.

\section{Sample Preparation and Experimental Techniques}
\label{sec:samples}

The fabrication of SIFS Josephson junctions with controllable and reproducible
properties was realized by the deposition of Nb/AlO$_x$/$\rm
Pd_{0.82}Ni_{0.18}$/Nb multilayers by UHV dc magnetron sputtering and
subsequent patterning of this multilayer stack using optical lithography, a
lift-off process, as well as reactive ion etching (RIE). Thermally oxidized
silicon wafers ($\sim 50$\,nm oxide thickness) were used as substrates.

In a first step the whole Nb/AlO$_x$/$\rm Pd_{0.82}Ni_{0.18}$/Nb (SIFS)
multilayer stack was sputter deposited in-situ in a UHV dc magnetron
sputtering system with a background pressure in the low $10^{-9}$\,mbar range.
This system is equipped with three sputter guns (Nb, $\rm Pd_{0.82}Ni_{0.18}$,
Al) and an Ar ion beam gun for surface cleaning. The sputtering chamber is
attached to a UHV cluster tool, allowing for the transfer of the samples to an
AFM/STM system for surface characterization without breaking vacuum. The
multilayer stack consists of a niobium base electrode with thickness $d_{\rm
1,Nb} = 85$\,nm, an aluminum layer of thickness $d_{\rm Al} = 4$\,nm, a
ferromagnetic $\rm Pd_{0.82}Ni_{0.18}$ (PdNi) interlayer with thickness $d_{\rm
F}$ ranging between 4 and 15\,nm, and finally a niobium top electrode with
thickness $d_{\rm 2,Nb} = 50$\,nm. The reproducible fabrication of
ferromagnetic SIFS Josephson junctions requires the precise control of the
thickness of the PdNi layer and the minimization of the roughness of the
involved interfaces. Therefore, we have carefully optimized the parameters of
the sputtering process (Ar pressure, power, substrate-target distance) for both
the Nb and PdNi layers to obtain films with very smooth surfaces. After
optimization, the rms roughness of the Nb base electrode was reduced to $\sim
0.4$\,nm.
% at a film thickness of 50\,nm.
The ferromagnetic PdNi layers showed a
slightly larger rms roughness of $\sim 0.8$\,nm. For the deposition of Nb and
Al an Ar pressure of $2.7\times 10^{-3}$\,mbar was used, resulting in a
deposition rate of 0.7\,nm/s at 200\,W for niobium and 0.2\,nm/s at 40\,W for
aluminum, respectively. PdNi was sputtered at a higher Ar pressure of $2\times
10^{-2}$\,mbar, resulting in a growth rate of 0.4\,nm/s at 40\,W. The
transition temperature of the Nb films was $T_{\rm c} = 9.2$\,K and the Curie
temperature of the PdNi films was determined to about 150\,K by SQUID
magnetometry.

A critical process step in the deposition of the Nb/AlO$_x$/$\rm
Pd_{0.82}Ni_{0.18}$/Nb (SIFS) multilayer stack is the fabrication of the
tunneling barrier. To achieve good reproducibility, the tunneling barrier was
realized by partial thermal oxidation of the 4\,nm thick Al layer inside the
sputtering chamber. The thickness of the AlO$_x$ tunneling barrier was adjusted
by varying the oxygen partial pressure and the duration of the thermal
oxidation process. The oxidation time was varied between 60 and 240\,min in a
pure oxygen atmosphere of 0.1\,mbar. We note that the thickness of the AlO$_x$
tunneling barrier determines the $R_{\rm n}\cdot A$ values of the junctions,
because the tunneling resistance is much larger than the resistance of the $\rm
Pd_{0.82}Ni_{0.18}$ layer.

\begin{figure}[tb]
  \centering
  \includegraphics[width=.95\columnwidth]{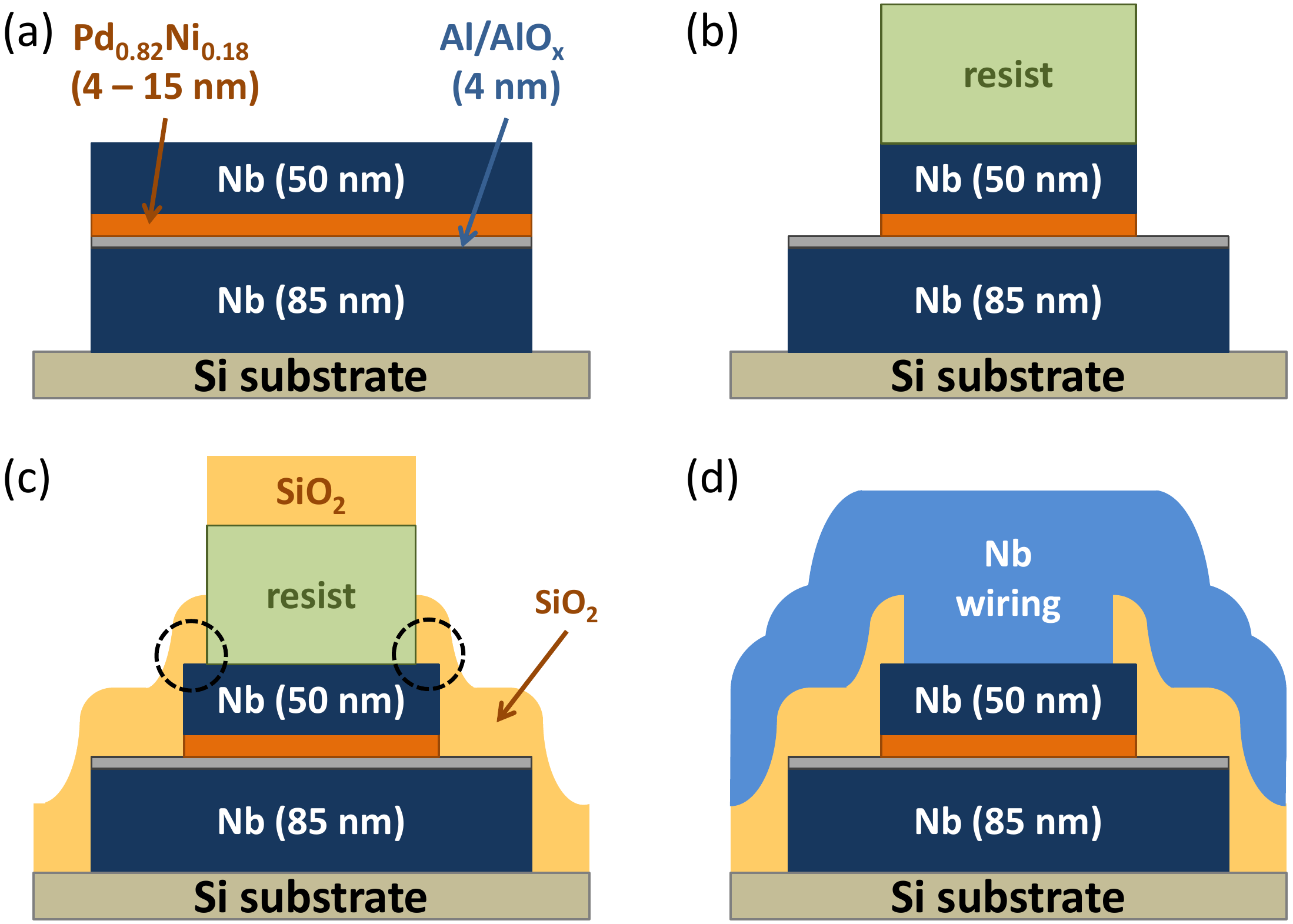}
  \caption{
Cross-sectional views of the SIFS Josephson junctions after different steps of
the fabrication process. (a) Cross-sectional view after the definition of the
about $100\,\mu$m wide strip forming the base electrode by a lift-off-process.
(b) Cross-sectional view after the etching of the mesa structure defining the
junction area by a RIE process. (c) Cross-sectional view after the additional
oxygen plasma process reducing the size of the resist stencil and the
subsequent deposition of the SiO$_2$ wiring insulation. The dashed circles highlight
the coverage of the edges of the niobium top electrode by the SiO$_2$ layer to
prevent shorts. (d) Cross-sectional view of the completed junction after the
deposition of the Nb wiring layer and patterning by a lift-off process.}
 \label{fig:Wild_EPJB2010_GInsituProc}
\end{figure}

After the controlled in-situ deposition of the Nb/AlO$_x$/$\rm
Pd_{0.82}Ni_{0.18}$/Nb multilayer stack, in the next step the SIFS Josephson
junctions are fabricated by a suitable patterning process. We used a
three-stage self-aligned process based on optical lithography, a lift-off
process and reactive ion etching (RIE). In the first step, the base electrode
is defined by patterning a long about $100\,\mu$m wide strip into the whole
multilayer stack. This is achieved by placing a suitable photoresist stencil on
the Si substrate and using a lift-off process after the deposition of the SIFS
multilayer stack. Fig.~\ref{fig:Wild_EPJB2010_GInsituProc}a shows a
cross-sectional view of the SIFS multilayer after this step. Next, the junction
area is patterned by etching a mesa structure into the SIFS stack by placing a
photoresist stencil on top of the SIFS stack. This resist stencil serves as the
etching mask in a RIE patterning process thereby defining the shape and size
of the junction area. Junction areas between $2.5\times 2.5\,\mu$m$^2$ and
$50\times 50\,\mu$m$^2$ have been realized. The RIE process was performed in a
SF$_6$ plasma (time: 70\,s, voltage: 300\,V). Note that the RIE process
selectively patterns the Nb top electrode and the PdNi layer because the
AlO$_x$ layer acts as an effective stopping layer. The resist stencil defining
the junction area is used for the lift-off process in the subsequent deposition
of the SiO$_2$ wiring insulation (self-aligned process). To avoid electrical
shorts between the wiring layer for the top electrode and the base electrode,
the lateral dimensions of the resist mask were reduced by about $150\,$nm using
an oxygen plasma process in the RIE system immediately after the mesa
patterning. In this way, the SiO$_2$ wiring insulation also covers the junction
edges, preventing electrical shorts (cf.
Fig.~\ref{fig:Wild_EPJB2010_GInsituProc}c). The 50\,nm thick SiO$_2$ wiring
insulation is deposited by rf-magnetron sputtering in a 75\%Ar/25\%O$_2$
atmosphere. In a last step, the 200\,nm thick niobium wiring layer is
deposited. The Nb deposition was done by dc magnetron sputtering and the
patterning was realized by optical lithography and a subsequent lift-off
process. To obtain a good superconducting contact between the Nb wiring layer
and the Nb top electrode, the surface of the top electrode has been cleaned
in-situ prior to the deposition of the wiring layer using an Ar ion gun. A
cross-sectional view of the completed junction is shown in
Fig.~\ref{fig:Wild_EPJB2010_GInsituProc}d. Of course, the junction fabrication
process allows for the fabrication of several junctions with different junction
areas on the same wafer. In this way the reproducibility of the process can be
checked by measuring the on-chip parameter spread.

The characterization of the junctions was performed in a $^3$He cryostat and a
$^3$He/$^4$He dilution refrigerator, allowing temperatures down to 300\,mK and
20\,mK, respectively. The dilution refrigerator was placed in a rf-shielded
room to reduce high-frequency noise. Furthermore, external magnetic fields have
been reduced by $\mu$-metal and/or cryoperm shields. Small magnetic fields
aligned parallel to the junction barrier could be applied by a superconducting
Helmholtz coil. The current-voltage characteristics (IVCs) have been measured
in a four-point configuration. A low-noise current source and voltage
preamplifier (Stanford Research SR~560) have been used. Both were battery
powered and placed inside the rf-shielded room. Derivatives of the IVCs have
been taken using a lock-in technique. The magnetic properties of the
ferromagnetic interlayer have been measured using a Quantum Design SQUID
magnetometer.

\section{Superconducting and Magnetic Properties}
\label{sec:properties}

In this section we discuss the superconducting and magnetic properties of the
SIFS Josephson junctions fabricated with the process described above. In
particular, we will address the magnetic properties of the $\rm
Pd_{0.82}Ni_{0.18}$ layer and the dependence of the product $V_{\rm c} = I_{\rm
c} R_{\rm n}$ of critical current $I_{\rm c}$ and normal resistance $R_{\rm n}$
of the junctions on the thickness $d_{\rm F}$ of the ferromagnetic interlayer.
We will see that this dependence shows a clear transition from $0$- to
$\pi$-coupled junctions at $d_{\rm F} \simeq 6$\,nm. In this article we will
focus mainly on the properties of the $\pi$-coupled SIFS junctions. We will
discuss their current-voltage characteristics (IVCs) and their magnetic field
dependence of the critical current, allowing us to determine the basic
parameters of the SIFS junctions.

\subsection{Magnetic Properties of the PdNi Layer}
\label{subsec:PdNi}

\begin{figure}[tb]
  \begin{center}
 \includegraphics[width=.95\columnwidth]{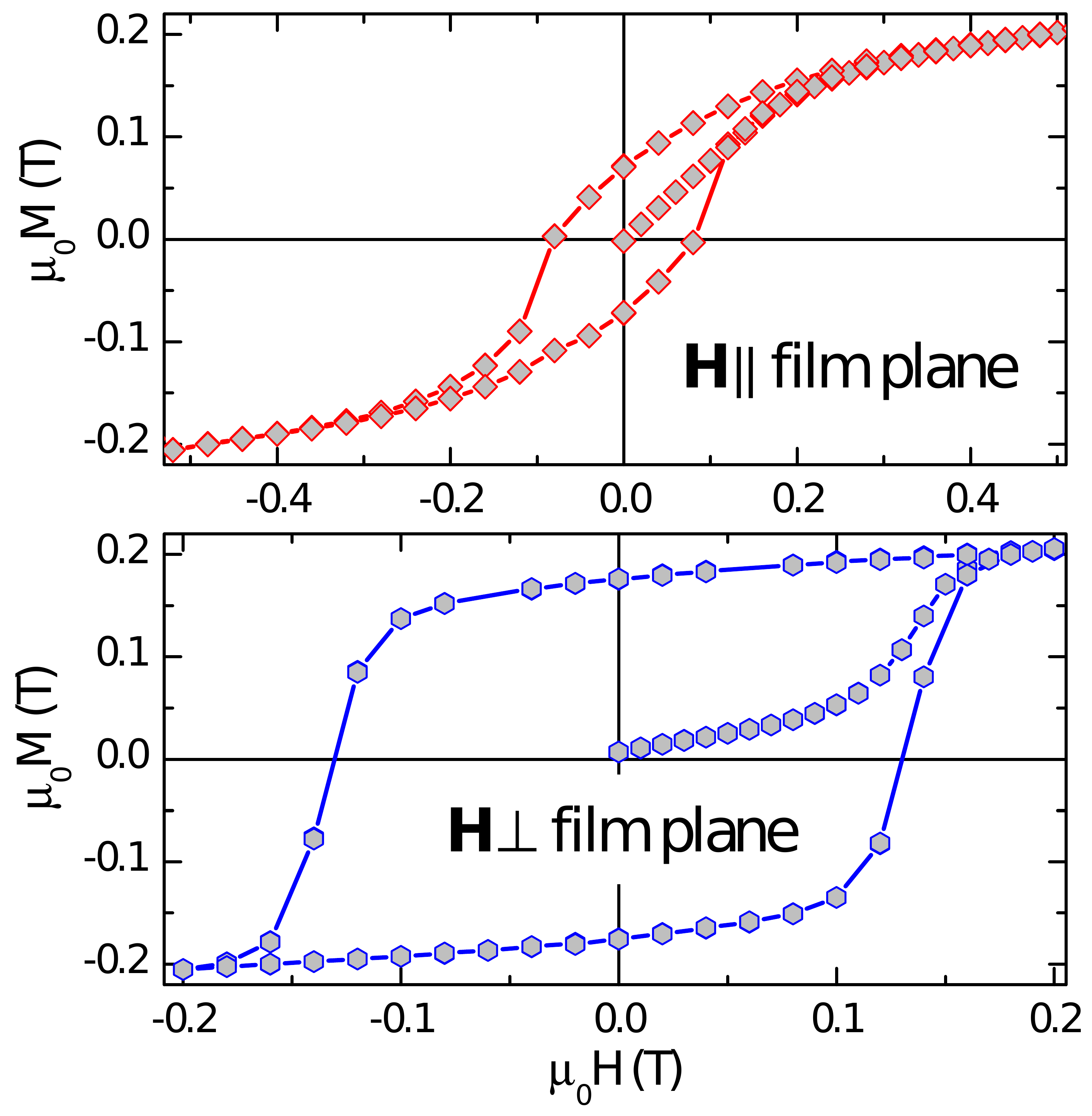}
  \caption{
Magnetization versus applied magnetic field curves of a 8.4\,nm thick $\rm
Pd_{0.82}Ni_{0.18}$ layer sandwiched in a Nb/AlO$_x$/PdNi/Nb multilayer stack
for the magnetic field applied parallel (top panel) and perpendicular (bottom
panel) to the film plane. The curves were measured at 11\,K, that is, just above
the critical temperature of the Nb films. The magnetization data have been
corrected by subtracting the diamagnetic background due to the metallic layers
and the substrate. This contribution is determined from the slope of the $M(H)$
curves at high fields above 1\,T where the magnetization of the ferromagnetic
layer is fully saturated. }
  \label{fig:Wild_EPJB2010_SQUID}
  \end{center}
\end{figure}

The Curie temperature of the $\rm Pd_{0.82}Ni_{0.18}$ layer has been determined
to about 150\,K by measuring the magnetization versus temperature dependence
using a Quantum Design SQUID magnetometer. For thin ferromagnetic films the
easy axis of the magnetization usually is parallel to the film plane to
minimize the free energy contribution due to the shape anisotropy. However, in
very thin films the magnetic surface energy may be dominant resulting in an
easy axis perpendicular to the film plane. To get information on the direction
of the easy axis we have recorded the magnetization versus applied magnetic
field curves of the thin $\rm Pd_{0.82}Ni_{0.18}$ layer within a
Nb/AlO$_x$/$\rm Pd_{0.82}Ni_{0.18}$/Nb multilayer stack with the magnetic field
applied in- and out-of-plane. The result is shown in
Fig.~\ref{fig:Wild_EPJB2010_SQUID}. Qualitatively, an almost rectangularly
shaped $M(H)$ curve is expected for the field applied along the easy axis
because the magnetization tries to stay along this preferred direction as long
as possible and than abruptly switches to the opposite direction at the
coercive field. In contrast, for the field applied along the hard axis the
magnetization is expected to be gradually rotated out of the easy axis
direction into the hard axis direction on increasing the applied magnetic
field, resulting in a gradually increasing and decreasing magnetization when
sweeping the field. As demonstrated by Fig.~\ref{fig:Wild_EPJB2010_SQUID}, the
measured $M(H)$ curve is much more rounded and has smaller remanent
magnetization, when the field is applied in-plane, whereas it has an almost
rectangular shape when the field is applied out-of-plane. This is clear
evidence for an out-of-plane anisotropy of the PdNi film. That is, the
out-of-plane (in-plane) direction is the magnetic easy (hard) axis in agreement
with literature~\cite{Khaire2009}. An out-of-plane anisotropy has been observed
also for CuNi~\cite{Ruotolo2004}. A more detailed analysis of the magnetic
anisotropy of the PdNi films would require additional experiments such as
ferromagnetic resonance and methods providing information on the domain
structure. However, it is difficult to perform such experiments with
ferromagnetic layers enclosed by two metallic Nb layers.

The saturation magnetization $M_{\rm s}$ of the  $\rm Pd_{0.82}Ni_{0.18}$ layer
is about $\mu_0M_{\rm s} \simeq 0.2$\,T in good agreement with literature
values~\cite{Khaire2009}. This magnetization corresponds to a magnetic moment
of slightly above $1\,\mu_{\rm B}$ per Ni atom, if one assumes that there is
negligible contribution of Pd. Here, $\mu_{\rm B}$ is Bohr's magneton. This is
close to the values in polycrystalline bulk samples, for which $\mu_{\rm Ni}
\simeq 1.1\,\mu_{\rm B}$ and $\mu_{\rm Pd} \simeq 0.1\,\mu_{\rm B}$ has been
reported~\cite{CablePRB1970}.  This suggests that there are no significant
magnetic dead layers. Since $M_{\rm s}$ was measured with an only 8\,nm thick
PdNi film, even very thin dead layers at the interfaces would result in a
significant reduction of $M_{\rm s}$. In the measurement of the magnetic field
dependence of the critical current of the SIFS junctions only an in-plane
magnetic field of the order of a few mT is applied. In this small field range
only the virgin $M(H)$ curve is relevant, which is about linear and almost
non-hysteretic. From the experimental data taken at 11\,K the slope of the
$M(H)$ curve is determined to $\mu = dM/dH = 1.8$.

\subsection{Crossover between 0- and $\pi$-Coupling}
\label{subsec:crossover}

\begin{figure}[tb]
  \centering
  \includegraphics[width=.95\columnwidth]{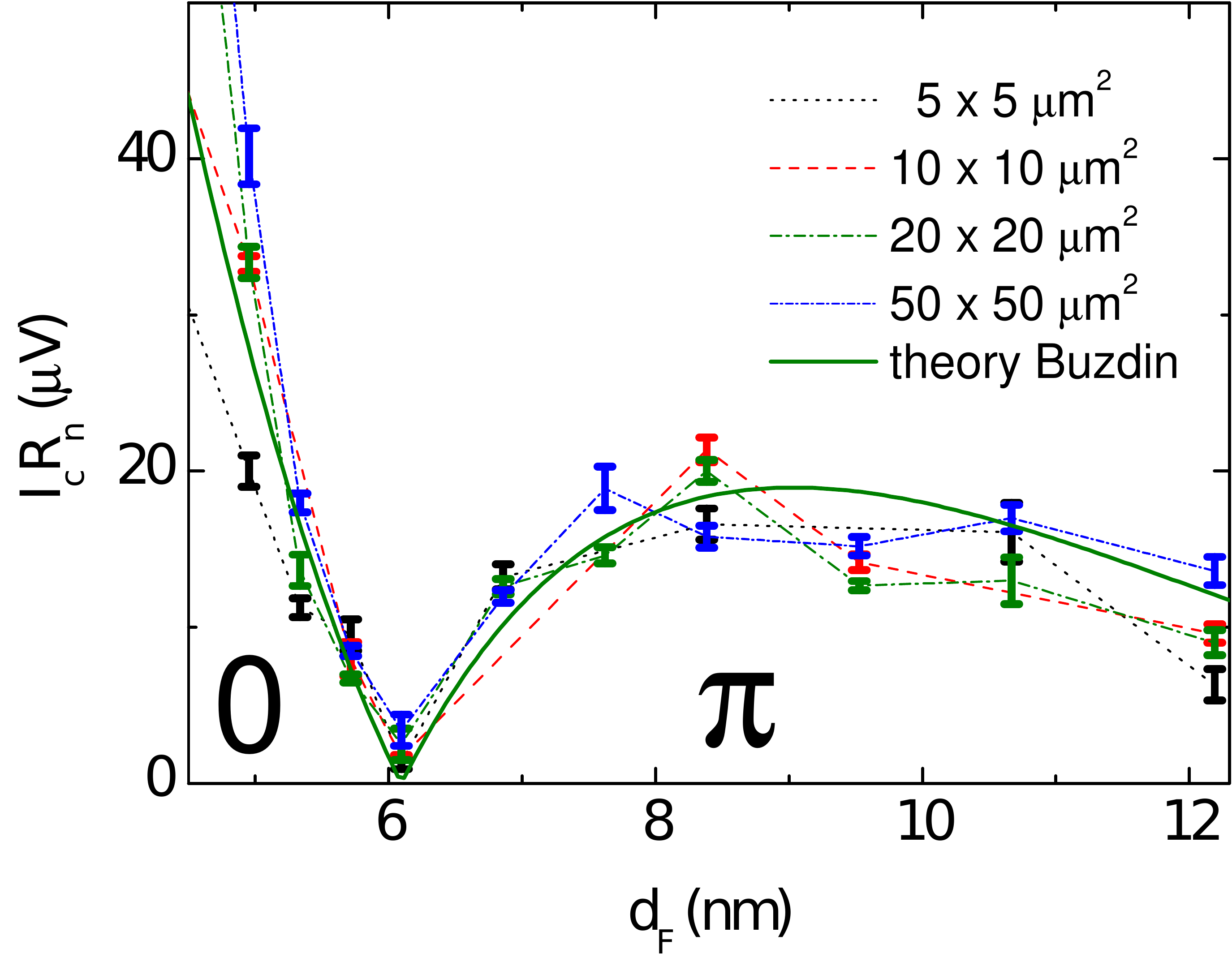}
  \caption{
Dependence of the $I_{\rm c}R_{\rm n}$-product of SIFS junctions on the
thickness $d_{\rm F}$ of the ferromagnetic $\rm Pd_{0.82}Ni_{0.18}$ layer for samples
with different junction areas. The broken lines are guides to the eye. The olive line
is a fit of the data to the theory of Buzdin {\em et
al.}~\cite{BuzdinPRB2003} using $\xi_{\rm F}=3.88\,$nm and $\pi \Delta_1\Delta_2/e
k_{\rm B}T_{\rm c} = 280\;\mu$V.}
  \label{fig:Wild_EPJB2010_IcD}
\end{figure}

We next discuss the dependence of the characteristic junction voltage $V_{\rm
c} = I_{\rm c}R_{\rm n}$ on the thickness $d_{\rm F}$ of the ferromagnetic
interlayer. The $V_{\rm c}$ values of a large number of junctions with
different $d_{\rm F}$ and junction areas $A$ ranging from $5\times
5\,\mu\mathrm{m}^2$ to $50\times 50\,\mu\mathrm{m}^2$ are plotted in
Fig.~\ref{fig:Wild_EPJB2010_IcD} versus the thickness $d_{\rm F}$ of the F
layer. To achieve this data, a series of junctions with different $d_{\rm F}$
was fabricated with otherwise identical parameters. In particular, all junctions
have AlO$_x$ barriers obtained by a 90\,min long thermal oxidation process,
resulting in $R_{\rm n} \cdot A \simeq 40\,\Omega\mu$m$^2$. This series clearly
shows a change of sign in the slope of the $I_{\rm c}R_{\rm n} (d_{\rm F})$
dependence at a $\rm Pd_{0.82}Ni_{0.18}$ layer thickness of $d_{\rm F} \simeq
6\,$nm. At this $d_{\rm F}$ value $I_{\rm c}$ approaches zero. This feature is
a clear signature of the crossover from $0$- to $\pi$-coupling on increasing
$d_{\rm F}$, corresponding to a change of sign of $I_{\rm c}$. Obviously, in
the experiment we can only measure the modulus of $I_{\rm c}$.

In order to theoretically describe the behavior shown in
Fig.~\ref{fig:Wild_EPJB2010_IcD} one has to distinguish different regimes
defined by three energy scales~\cite{BuzdinRMP2004}. These scales are the
exchange energy $E_{\rm ex}$ in the ferromagnet, the energy gap $\Delta$ of the
superconductor, and $\hbar/\tau$, where $\tau$ is the elastic scattering time
in the ferromagnet. In our samples, $E_{\rm ex} \gg \Delta$ ($\Delta =
1.5$\,meV for Nb) as in the overwhelming part of other reports on SFS or SIFS
junctions. However, there is a significant variation in $\hbar/\tau$ relative
to the other two energy scales. The true clean-limit holds for $E_{\rm ex},
\Delta \gg \hbar/\tau$. In this case the mean free path $\ell = v_{\rm F} \tau$
is large compared to the clean-limit superconducting coherence length
$\widetilde{\xi}_{\rm s} = \hbar v_{\rm F}/\pi\Delta$ and exchange length
$\widetilde{\xi}_{\rm F} = \hbar v_{\rm F}/2E_{\rm ex}$. Here, $v_{\rm F}$ is
the Fermi velocity in the respective material. The true dirty limit holds for
$E_{\rm ex}, \Delta \ll \hbar/\tau$. In this case the mean free path $\ell =
v_{\rm F} \tau$ is the smallest length scale and the dirty-limit
superconducting coherence length $\xi_{\rm s} \simeq \sqrt{\widetilde{\xi}_{\rm
s}\ell} \simeq \sqrt{\hbar D/\pi\Delta}$ and exchange length $\xi_{\rm F}
\simeq \sqrt{\widetilde{\xi}_{\rm F}\ell} \simeq \sqrt{\hbar D/E_{\rm ex}}$ are
given by the geometric mean. Here, $D$ is the diffusion coefficient in the
respective material. There is also an intermediate regime, where $E_{\rm ex} >
\hbar/\tau$ and $ \Delta < \hbar/\tau$, which is the most complicated
situation. This regime may particularly apply for a ferromagnet with large
$E_{\rm ex}$.

Due to the small mean free path in the PdNi alloy and Nb films, for our samples
the simple dirty limit holds. Several theoretical models have been proposed for
this
limit~\cite{BuzdinRMP2004,KontosPRL2002,DemlerPRB1997,BuzdinPRB2003,BergeretPRB2003,BergeretPRB2007,FaurePRB2006}.
For weak ferromagnets such as PdNi the spin-up and spin-down subbands can be
treated identically (same Fermi velocity and mean free path) resulting in a
single characteristic length scale
\begin{eqnarray}
\xi_{\rm F} & = & \sqrt{\frac{\hbar D}{E_{\rm ex}}}
\label{eq:xiF}
\end{eqnarray}
for the decay and oscillation of the critical current as a function of $d_{\rm
F}$. We note, however, that in the presence of spin-flip or spin-orbit
scattering the decay and oscillation of $I_{\rm c}$ is governed by two
different length
scales~\cite{DemlerPRB1997,BergeretPRB2003,BergeretPRB2007,FaurePRB2006}.

The solid olive line in Fig.~\ref{fig:Wild_EPJB2010_IcD} is obtained by fitting
the data using the simple dirty limit expression of Buzdin {\em et
  al.}~\cite{BuzdinPRB2003}. The fit yields $\xi_{\textrm{F}}=3.88\,$nm and
$\frac{\pi\Delta_1\Delta_2}{e k_{\rm B}T_{\rm c}} = 280\,\mu$V. Here, $T_{\rm c}
= 9.2$\,K is the critical temperature of the niobium layers and $\Delta_1$ and
$\Delta_2$ are the superconducting order parameters just at the boundary with
the ferromagnetic layer. They are certainly much smaller than the
superconducting gap in bulk niobium but difficult to be determined in the
geometry of our experiment. The value $\xi_{\textrm{F}}=3.88\,$nm found for our
SIFS junctions agrees very well with literature
values~\cite{KontosPRL2002,BuzdinPRB2003,KontosPRL2001}. The value of
$280\,\mu$V found for $\frac{\pi\Delta_1\Delta_2}{e k_{\rm B}T_{\rm c}}$ for our
SIFS junctions is larger than the value of $110\,\mu$V obtained by Kontos and
coworkers~\cite{KontosPRL2002,BuzdinPRB2003}. Unfortunately, the detailed
comparison of different experiments is difficult because the authors often do
not state whether they are plotting $I_{\rm c}R_{\rm n}(d_{\rm F})$ or $I_{\rm
  c}R_{\rm sg}(d_{\rm F})$. If we use the much higher subgap resistance $R_{\rm
  sg}$, the corresponding value derived for $\frac{\pi\Delta_1\Delta_2}{e k_{\rm
    B}T_{\rm c}}$ would be about ten times larger. We finally note that the
measured $I_{\rm c}(d_{\rm F})$ dependence can be well explained by dirty limit
theory with a single length scale $\xi_{\rm F}$, suggesting that spin-flip or
spin-orbit scattering do not play a dominant role.

Beyond the parameter $\xi_{\rm F}$ describing the characteristic length of
superconducting correlations in the F layer, the dimensionless parameter
\begin{eqnarray}
\gamma_{\rm B} & = & \frac{\rho_{\rm B} \sigma_{\rm F}}{\xi_{\rm F}}
\label{eq:gammaB}
\end{eqnarray}
is used to describe ferromagnetic Josephson junctions. It characterizes the
transparency of the F/S interfaces with $\rho_{\rm B}$ the interface resistance
times area and $\sigma_{\rm F}$ the conductivity of the F layer. In general,
two $\gamma_{\rm B}$ values for the two F/S interfaces have to be used. In our
experiment the presence of the additional AlO$_x$ barrier at one S/F boundary
can be modeled by a very low transparency interface ($\gamma_{\rm B1} \gg 1$),
while the other boundary has high transparency ($\gamma_{\rm B2} \ll 1$) for
the in-situ fabricated stacks, i.e. $\gamma_{\rm B} \simeq \gamma_{\rm B1}$. We
further note that in general the measured total $R_{\rm n} \cdot A$ product can
be expressed as
\begin{eqnarray}
R_{\rm n} \cdot A & = & \rho_{\rm tun} + \rho_{\rm int} + \frac{d_{\rm F}}{\sigma_{\rm F}} \;\; .
\label{eq:Rn}
\end{eqnarray}
Here, $\rho_{\rm tun}$ and $\rho_{\rm int}$ are the resistance times area
values due to the tunneling barrier and the PdNi/Nb interfaces. For our
junctions, $\rho_{\rm tun} \gg \rho_{\rm int}$ and, moreover, $\rho_{\rm tun}
\gg d_{\rm F} / \sigma_{\rm F}$. With $\sigma_{\rm PdNi} \simeq 10^7
\Omega^{-1}\mathrm{m}^{-1}$ we estimate $d_{\rm F} / \sigma_{\rm F} \simeq
10^{-3}\Omega \mu\mathrm{m}^2$ which is by about five orders of magnitude
smaller than the measured $R_{\rm n}\cdot A$ values. That is, the contribution
of the F layer to $R_{\rm n}\cdot A$ is negligible. Hence, $R_{\rm n}\cdot A
\simeq \rho_{\rm tun}$, meaning that the measured $R_{\rm n}\cdot A$ values are
dominated by the AlO$_x$ tunneling barrier as expected. In this case we can
write $\gamma_{\rm B} \simeq R_{\rm n} A \, \sigma_{\rm F} /\xi_{\rm F}$. With
$\sigma_{\rm PdNi} \simeq 10^7 \Omega^{-1}\mathrm{m}^{-1}$, $\xi_{\rm F} =
3.88$\,nm and the measured $R_{\rm n}\cdot A$ value of about
$40\,\Omega\mu\mathrm{m}^2$ for this junction series we estimate $\gamma_{\rm
B} \simeq 10^5$. This high value is not surprising due to the additional
tunneling barrier in our junctions.

The derived $\xi_{\rm F}$ values can be used to estimate the exchange energy in
PdNi. Using a Fermi velocity $v_{\rm F} \simeq 5\times
10^5$m/s~\cite{DyePRB1981} and the fact that the mean free path $\ell$ for the
very thin PdNi layers is about given by the film thickness $d_{\rm F}$, we
obtain $D \simeq v_{\rm F} \ell \simeq v_{\rm F}d_{\rm F}= 5\times
10^{-3}$m$^2$/s for $d_{\rm F}= 10$\,nm. With this value we derive $E_{\rm ex}
\simeq \hbar D/\xi_{\rm F}^2 \simeq 20$\,meV using $\xi_{\rm F} = 3.88$\,nm.
This value is in good agreement with values between about 10 and 50\,meV quoted
in previous work~\cite{KontosPRL2002,Bauer2004,Khaire2009,CirilloPRB2005}. With
the same numbers we obtain $\hbar / \tau \simeq \hbar v_{\rm F}/d_{\rm F}
\simeq 30$\,meV. This shows that the SIFS junctions with PdNi interlayer are
close to the intermediate regime since $E_{\rm ex} \sim \hbar/\tau$.

\subsection{IVCs and Magnetic Field Dependence of the Critical Current}
\label{subsec:IVC_IcH}

\begin{figure}[tb]
  \begin{center}
 \includegraphics[width=.99\columnwidth]{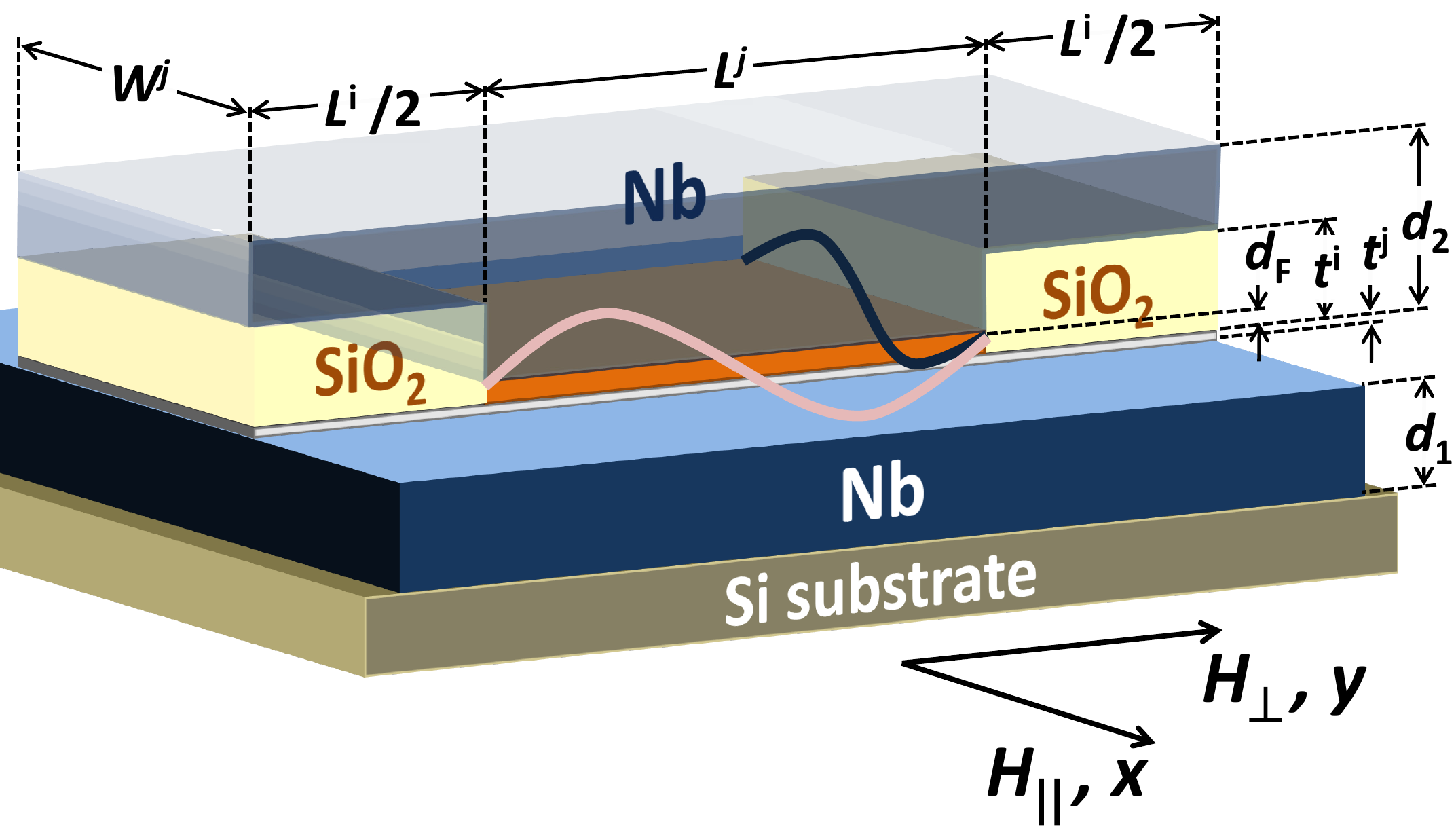}
  \caption{
Sketch of the SIFS junction geometry. The junction area of length $L^{\rm j}$
and width $W^{\rm j}$ is formed by the junction electrodes of thickness $d_1$
and $d_2$ separated by the tunneling barrier with thickness $t^{\rm j}$ and
dielectric constant $\epsilon^{\rm j}$ and the ferromagnetic layer with
thickness $d_{\rm F}$. Next to the junction area there is an overlap of the Nb
wiring layer and the base electrode, forming the so-called idle region of
length $L^{\rm i}/2$ at each side ($L^{\rm i}=$ a few $10\,\mu\mathrm{m}$ and $W^{\rm i}
\simeq 0$ in our junctions). Both layers are separated by the SiO$_2$ wiring
insulation of thickness $t^{\rm i}$ with dielectric constant $\epsilon^{\rm i}$.
}
  \label{fig:Overlap}
  \end{center}
\end{figure}

In the following we discuss the IVCs and magnetic field dependence of the
critical current. We focus on SIFS junctions with a F layer thickness of
$d_{\rm F}=8.4\,$nm resulting in $\pi$-coupling. The AlO$_x$ tunneling barrier
has been achieved by thermal oxidation for 4\,h in pure oxygen to obtain high
$R_{\rm n}\cdot A$ values and, in turn, low damping of the self excited
resonances. We start our discussion by considering the effect of the specific
junction geometry (cf. Fig.~\ref{fig:Overlap}) used in our experiments on the
derived junction parameters. Although these effects are often neglected in
literature, it is mandatory to take them into account in a detailed evaluation.
Of particular importance are the finite thickness of the junction electrodes
and the so-called idle region. The latter is formed when the wiring of the top
electrode is deposited to complete the junction structure. Then the bottom
electrode, the SiO$_2$ wiring insulation and the wiring layer are forming an
SIS structure next to the junction area.

\paragraph{Effective Magnetic Thickness}

It is well known that a magnetic field applied parallel to the surface of a
bulk superconductor decays exponentially inside the superconductor due to the
Mei{\ss}ner effect~\cite{Tinkham}. The characteristic screening length is the
London penetration depth $\lambda_{\rm L}$ which is about 90\,nm for
Nb~\cite{Kohlstedt1995}. However, in our junctions we are using electrodes of
finite thickness $d$. In such a thin film superconductor the magnetic field
dependence perpendicular to the film ($z$-direction) is obtained to
\begin{eqnarray}
H(z) & = & \frac{H_{\rm ext,1} + H_{\rm ext,2}}{2} \; \frac{\cosh (z/\lambda_{\rm L})}{\cosh (d/2\lambda_{\rm L})}
 \nonumber \\
        & &  - \frac{H_{\rm ext,1} - H_{\rm ext,2}}{2} \; \frac{\sinh (z/\lambda_{\rm L})}{\sinh (d/2\lambda_{\rm L})} \; .
\label{eq:Hzthinfilm}
\end{eqnarray}
by solving the London equations~\cite{Tinkham}. Here, the film is assumed to
extend from $-d/2$ to $+d/2$, and $H_{\rm ext,1}$ and $H_{\rm ext,2}$ are the
external magnetic fields applied parallel to the film at both sides. The
boundary conditions and the resulting current distributions are different for
different physical situations. If the field is confined on one side of the
superconducting film, we have $H_{\rm ext,1} =H_0$ and $H_{\rm ext,2} =0$. For
Josephson junctions this situation applies to cases where one is considering
only the fields related to supercurrents flowing in the junction electrodes. If
$d<\lambda_{\rm L}$, screening currents are confined to a length scale smaller
than $\lambda_{\rm L}$, resulting in an enhanced kinetic inductance $L_{\rm s}
= \mu_0\lambda_{\rm L} \coth (d/\lambda_{\rm L})$ as compared to a bulk
superconductor with $L_{\rm s} = \mu_0\lambda_{\rm L}$~\cite{Swihart61}. That
is, the thin film superconductor behaves equivalent to the bulk one with an
effective screening length $\lambda_{\rm L} \coth (d/\lambda_{\rm L})$. Then,
the magnetic penetration in the barrier layer and the junction electrodes can
be described by the effective magnetic thickness
\begin{eqnarray}
t^{\rm j}_{\rm B} & = & t^{\rm j} + \mu d_{\rm F}
                          +\lambda_{\rm L1}\coth \frac{d_1}{\lambda_{\rm L1}}
                          +\lambda_{\rm L2}\coth \frac{d_2}{\lambda_{\rm L2}} \; .
\label{eq:effectivethickness1}
\end{eqnarray}
Here, $t^{\rm j}$ is the thickness of the oxide barrier, $d_{\rm F}$ the
thickness of the F layer, $d_1$ and $d_2$ are the thicknesses of the Nb
junction electrodes, and $\lambda_{\rm L1}$ and $\lambda_{\rm L2}$ the
corresponding penetration depths. For our junctions we have $t^{\rm j}=4$\,nm
(we assume for simplicity that the oxide thickness is equal to the thickness of
the Al layer), $d_1=85$\,nm, $d_2=250$\,nm, $\lambda_{\rm L1} =\lambda_{\rm
L2}=\lambda_{\rm Nb}=90$\,nm~\cite{Kohlstedt1995}, $d_{\rm F} = 8.4$\,nm and
$\mu = 1.8$ resulting in $t_{\rm B}\simeq 230$\,nm. For the idle region next to
the junction area (cf. Fig.~\ref{fig:Overlap}) an expression equivalent to
eq.(\ref{eq:effectivethickness1}) is obtained by replacing $t^{\rm j}$ by
$t^{\rm i}$ on the right hand side and setting $d_{\rm F}=0$.

Another interesting case is the situation where the same magnetic field is
applied at both sides of the superconducting film. Here, $H_{\rm ext,1} =
H_{\rm ext,2} =H_0$ and hence $H(z) = H_0 \cosh (z/\lambda_{\rm L})/\cosh
(d/2\lambda_{\rm L})$. This case applies when we are measuring the dependence
of the junction critical current on a magnetic field applied parallel to the
junction electrodes. Since the junction (we are considering only short
junctions here) cannot screen the applied field from the region between the
junction electrodes, the same field is present at both surfaces of the junction
electrodes. To derive an effective magnetic thickness for this situation we
consider the total flux threading the junction. The latter is obtained by
integrating eq.(\ref{eq:Hzthinfilm}) along the $z$-direction. Whereas $\int
H(z) dz = 2H_0\lambda_{\rm L}$ for bulk electrodes ($d\gg \lambda_{\rm L})$,
for thin film electrodes we obtain $\int H(z) dz = 2H_0\lambda_{\rm L}\tanh
(d/2\lambda_{\rm L})$. That is, regarding the flux the thin film superconductor
behaves equivalent to the bulk one with an effective screening length
$\lambda_{\rm L} \tanh (d/2\lambda_{\rm L})$. Then, for the description of the
total flux threading the junction we can use the effective magnetic
thickness~\cite{Weihnacht1969}
\begin{eqnarray}
\widetilde{t}^{\rm j}_{\rm B} & = & t^{\rm j} + \mu d_{\rm F}
                                    +\lambda_{\rm L1}\tanh \frac{d_1}{2\lambda_{\rm L1}}
                                    +\lambda_{\rm L2}\tanh \frac{d_2}{2\lambda_{\rm L2}} \; .
\label{eq:effectivethickness2}
\end{eqnarray}

\paragraph{Effective Swihart Velocity and Frequencies of Resonant Modes}

The idle region shown in Fig.~\ref{fig:Overlap} influences the dynamic
properties of the junction. The reason is that the idle region acts as a
dispersive transmission line parallel to the junction edges. As discussed in
detail in section~\ref{sec:Fiske}, the junction forms a transmission line
resonator supporting so-called Fiske resonances~\cite{PhysRevFiske}. For a
junction without any idle region these resonances appear at frequencies
$\omega_n/2\pi = n (\bar{c}/2L^{\rm j})$ corresponding to voltages
\begin{eqnarray}
V_n & = &  n \, \frac{\Phi_0 \bar c}{2 L^{\rm j}} \; .
\label{eq:Fiske5}
\end{eqnarray}
Here $n = 1,2,3, \ldots$ and
\begin{eqnarray}
\bar{c} & = & c \; \sqrt{\frac{t^{\rm j}}{\epsilon^{\rm j} \; t_{\rm B}^{\rm j}}}
 \;
\label{eq:Swihart1}
\end{eqnarray}
is the Swihart velocity~\cite{Swihart61} with $c$ the velocity of light in
vacuum. Evidently, $\bar{c} \ll c$ as the junction barrier thickness $t^{\rm
j}$ is by almost two orders of magnitude smaller than the effective magnetic
thickness $t_{\rm B}^{\rm j}$ of the junction.

For a junction with idle region, the resonance frequencies and corresponding
voltages are shifted. We first discuss the effect of a lateral idle region
extending parallel to the resonant mode. In this case we have to consider the
junction transmission line in parallel with the two transmission lines of the
idle regions at both sides of the junction. To each isolated transmission line
we can assign the inductance $\mathcal{L}^{\rm i,j}=\mu_0 t_{\rm B}^{\rm
i,j}/W^{\rm i,j}$ and capacitance $\mathcal{C}^{\rm
i,j}=\epsilon_0\epsilon^{\rm i,j} W^{\rm i,j} /t^{\rm i,j}$ each per unit
length, where the indices $i$ and $j$ refer to the idle and junction region.
The corresponding phase velocities are $v^{\rm i,j}_{\rm ph} =
\sqrt{\mathcal{L}^{\rm i,j}\mathcal{C}^{\rm i,j}}$ with $v^{\rm j}_{\rm ph} =
\bar{c}$. The phase velocity of the combined transmission lines is given by
$v_{\rm ph} = 1/\sqrt{\mathcal{L} \mathcal{C} }$ with $\mathcal{L}^{-1} =
(\mathcal{L}^{\rm i})^{-1} + (\mathcal{L}^{\rm j})^{-1}$ and $\mathcal{C} =
\mathcal{C}^{\rm i} + \mathcal{C}^{\rm j}$. Evidently, the idle region increases
the capacitance and decreases the inductance per unit length. However, the
inductance effect is dominant since usually $t_{\rm B}^{\rm j}/t_{\rm B}^{\rm
i} \gg t^{\rm j}/t^{\rm i}$. This is true also for our samples. Therefore, the
phase velocity is increased by the idle region. A more detailed analysis
yields~\cite{Lee91,IEEEApplSup2Lee,Monaco1995}
\begin{eqnarray}
v_{\rm ph} & = & \bar{c} \; \sqrt{\frac{1+ \frac{t_{\rm B}^{\rm j}}{t_{\rm B}^{\rm i}}\;
\frac{W^{\rm i}}{W^{\rm j}} }{ 1 + \frac{t^{\rm j} \epsilon^{\rm i}}{t^{\rm i}\epsilon^{\rm j}}
\; \frac{W^{\rm i}}{W^{\rm j}} }}
 \; .
\label{eq:Swihart10}
\end{eqnarray}

We next discuss the effect of a longitudinal idle region extending
perpendicular to the resonant mode. In this situation the idle region can be
considered as a lumped capacitance loading the junction at its ends, since the
wave length usually is much larger than the dimension of the idle region. For
short junctions ($L^{\rm j}, W^{\rm j} \ll \lambda_{\rm J}$) detailed
calculations yield~\cite{Monaco1995}
\begin{eqnarray}
v_{\rm ph} & = & \bar{c} \; \frac{1}{1+4\left( \sqrt{\frac{C_1}{C_0} +1} - 1\right) }
 \;.
\label{eq:Swihart11}
\end{eqnarray}
Here, $C_1$ is the lumped load capacitance and $C_0$ the total junction
capacitance. We see that the effect of $C_1$ is to decrease the phase velocity.
However, since usually $t^{\rm i}/\epsilon^{\rm i} \gg t^{\rm j}/\epsilon^{\rm
j}$ we have $C_1 \ll C_0$, that is, the effect is quite small even if the idle
region has similar area as the junction. With $t^{\rm i}=50$\,nm,
$\epsilon^{\rm i} \simeq 3.9$, $t^{\rm j} = 2$\,nm, $\epsilon^{\rm j} \simeq
9.1$, $t_{\rm B}^{\rm j} \simeq 230$\,nm, $t_{\rm B}^{\rm i} \simeq 260$\,nm
and $W^{\rm i}/W^{\rm j} =0.8$ (typical for a $50\times 50\,\mu\mathrm{m}^2$
junction) we estimate $v_{\rm ph} \simeq 1.22\,\bar{c}$ for the lateral and
$v_{\rm ph} \simeq 0.95\,\bar{c}$ for the longitudinal mode.

\paragraph{Current-Voltage Characteristics}

\begin{figure}[tb]
  \centering
  \includegraphics[width=.95\columnwidth]{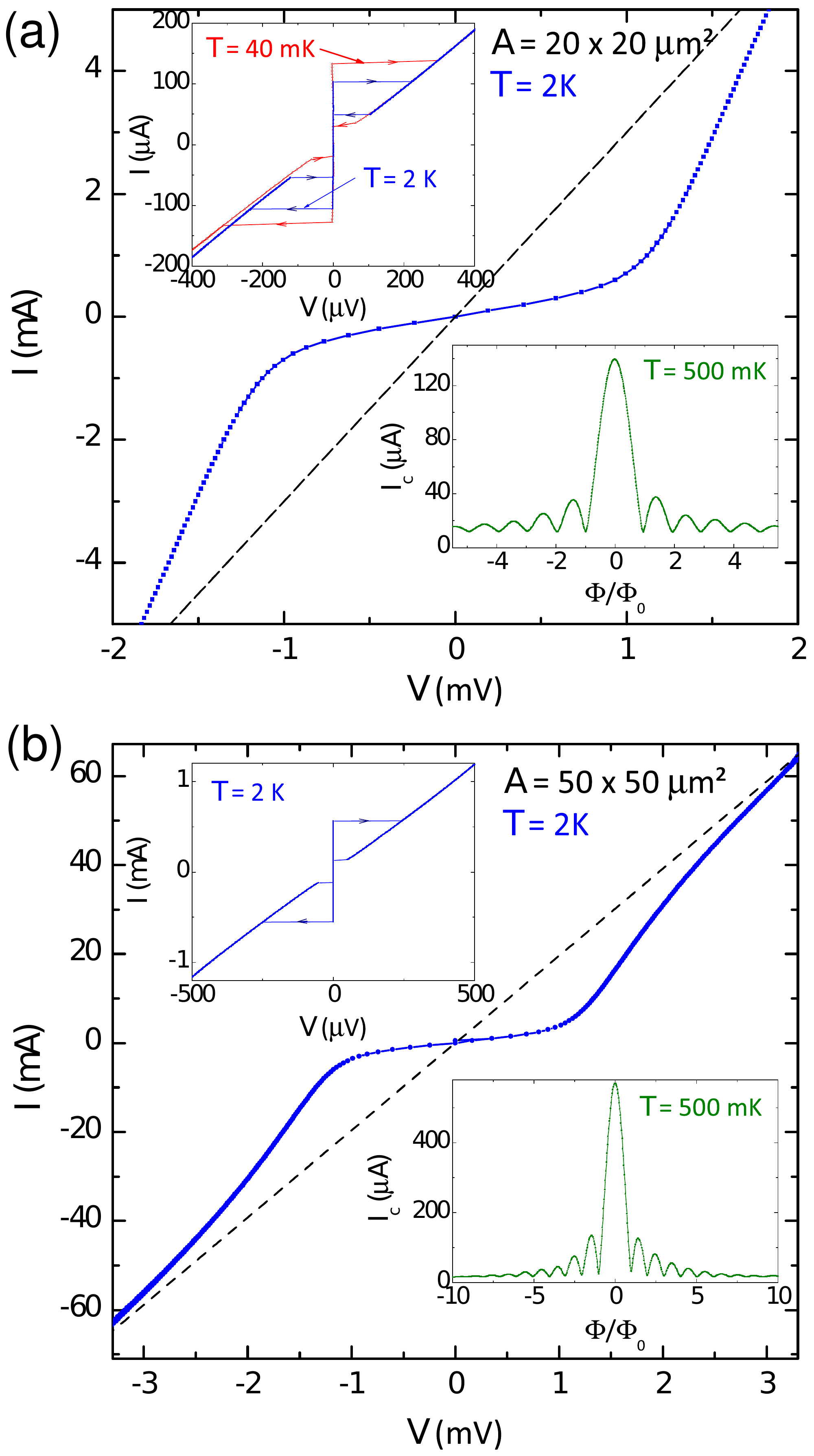}
  \caption{
IVCs obtained for $\pi$-coupled SIFS junctions with $A=20\times 20\,\mu$m$^2$ (a) and $50\times
50\,\mu$m$^2$ (b). The dashed lines indicate the ohmic behavior $V=IR_{\rm n}$
approached at large voltages. The upper and lower insets show enlarged views of the IVCs
around zero voltage and the magnetic field dependence of $I_{\rm c}$,
respectively. The magnetic field is applied parallel to the junction barrier.
The total magnetic flux $\Phi=\Phi_{\rm H}+\Phi_{\rm
M}$ threading the junctions originates from the applied magnetic field
($\Phi_{\rm H}$) and the magnetization of the F interlayer
($\Phi_{\rm M}$). The IVCs are measured at $2\,$K, while the $I_{\rm c}(\Phi)$
dependencies are taken at $500\,$mK. The red IVCs in the top inset of (a) is
measured at $40\,$mK.}
  \label{fig:Wild_EPJB2010_IVCs2050}
\end{figure}

Figure~\ref{fig:Wild_EPJB2010_IVCs2050} shows the IVCs of two $\pi$-coupled
SIFS junctions with junction areas $A=20 \times 20\,\mu$m$^2$ and $50\times
50\,\mu$m$^2$. From the IVCs we can determine several relevant junction
parameters. First, the current value for the switching from the zero to the
finite voltage state and vice versa gives the critical current $I_{\rm c}$ and
the so-called retrapping current $I_{\rm r}$ (see insets of
Fig.~\ref{fig:Wild_EPJB2010_IVCs2050}). The corresponding current densities
$J_{\rm c}$ and $J_{\rm r}$ are obtained by dividing by the junction area $A$.
From $J_{\rm c}$ together with the effective magnetic thickness $t_{\rm B}^{\rm
j}$ of the junction (cf. eq.(\ref{eq:effectivethickness1})) the Josephson
penetration depth $\lambda_{\rm J} = \sqrt{\hbar /2e \mu_0 t_{\rm B}^{\rm j}
J_{\rm c}}$ is derived. At $T=2$\,K we obtain $\lambda_{\rm J} \simeq
70\,\mu\mathrm{m}$. Since this value is larger than the lateral junction
dimensions, we are dealing with small Josephson junctions. Using the simple
resistively and capacitively shunted junction (RCSJ) model~\cite{APLstewart},
we can further derive the junction quality factor $Q_{\rm IVC} =
\sqrt{\beta_{\rm C}}=\omega_{\rm p}/\omega_{\rm RC}= 4I_{\rm c}/\pi I_{\rm r}$.
Here, $\beta_{\rm C}$ is the Stewart-McCumber parameter, $\omega_{\rm p}$ the
junction plasma frequency, and $1/\omega_{\rm RC}$ the $RC$ time constant of
the junction. Second, from the asymptotic behavior at voltages large compared
to the gap sum voltage the normal resistance $R_{\rm n}$ and the normal
resistance times area product, $R_{\rm n}\cdot A$, are obtained which are about
temperature independent. The $R_{\rm n}$ value has to be distinguished from the
temperature dependent subgap resistance $R_{\rm sg}$ obtained from the slope of
the IVCs at voltages well below the gap sum voltage. For SIFS junctions,
$R_{\rm sg}$ is expected to increase with decreasing temperature in agreement
with the experimental data. At 2\,K, for our SIFS junctions $R_{\rm sg}$ is
almost an order of magnitude larger than $R_{\rm n}$.

We note that the values of $I_{\rm c}$ and $I_{\rm r}$ may be reduced and
enhanced, respectively, by premature switching due to thermal activation or
external high-frequency noise. This is particularly true for small area
junctions with small absolute values of $I_{\rm c}$ and $I_{\rm r}$. In turn,
this results in reduced values of the quality factor. This effect is shown in
the inset of Fig.~\ref{fig:Wild_EPJB2010_IVCs2050}a, where the IVCs of a $20
\times 20\,\mu$m$^2$ junction are shown for $T=2$\,K and 40\,mK. The 40\,mK
data were taken in a well shielded dilution refrigerator using various filters
at different temperature stages in the current and voltage lines, including
stainless steel powder filters. Clearly, at 40\,mK significantly larger $I_{\rm
c}$ and smaller $I_{\rm r}$ values are observed resulting in an about three
times larger quality factor $Q_{\rm IVC}=6.2$. These enhanced/reduced values
are only partly caused by lowering the temperature but mostly by the reduced
thermal and external noise. For large area junctions this effect is negligible.
Here, $I_{\rm c}$ and $I_{\rm r}$ are large, making the relative effect of the
equivalent noise current very small. For example, at $T=2$\,K $Q_{\rm IVC}=5.7$
is obtained for the $50 \times 50\,\mu$m$^2$ junction, whereas $Q_{\rm IVC} <
2$ for the $20 \times 20\,\mu$m$^2$ junction fabricated on the same chip. This
small quality factor is related to the smaller $I_{\rm c}$ and $I_{\rm r}$
values of the small area junction, making it more susceptible to thermal and
external noise. In Table~\ref{tab:JJParams} we have tabulated the junction
parameters derived from the IVCs of the $50 \times 50\,\mu$m$^2$ and $20 \times
20\,\mu$m$^2$ junction at 2\,K and 40\,mK, respectively.

\begin{table}[tb]
\caption{Critical current and resistance values obtained from the IVCs of the
SIFS junctions with $d_{\rm F}=8.4$\,nm shown in
Fig.~\ref{fig:Wild_EPJB2010_IVCs2050} and parameters derived from them. The
quantities are listed for a $50 \times 50\,\mu$m$^2$ and $20 \times
20\,\mu$m$^2$ junction at 2\,K and 40\,mK, respectively.}
  \label{tab:JJParams}
  \begin{center}
  \begin{tabular}{lcc}
   \hline\noalign{\smallskip}
    $A$ ($\mu$m$^2$)  \hspace*{20mm}        & $20\times 20$ & $50\times 50$ \\
    $T$ (K)                                 & 40\,mK & 2K \\
   \noalign{\smallskip}\hline\noalign{\smallskip}
    $I_{\rm c}$ ($\mu$A)        & 131 & 555 \\
    $J_{\rm c}$ (A/cm$^2$)      & 33 & 22 \\
    $I_{\rm r}$ ($\mu$A)        & 27 &  125\\
    $R_{\rm n}$ ($\Omega$)      & 0.33 & 0.051\\
    $R_{\rm n}\cdot A$ ($\Omega\mu$m$^2$)   & 133 & 128 \\
    $R_{\rm sg}$ ($\Omega$)                 & 2.14 & 0.44 \\
    $I_{\rm c}R_{\rm n}$ ($\mu$V)           & 44 & 28.3 \\
    $I_{\rm c}R_{\rm sg}$ ($\mu$V)          & 280 & 244 \\
    $\lambda_{\rm J} (\mu$m)                & 59  & 71\\
    $Q_{\rm IVC}$                           & 6.2 & 5.7 \\
   \noalign{\smallskip}\hline
  \end{tabular}
  \end{center}
\end{table}

\paragraph{Magnetic Field Dependence of the Critical Current}

The bottom insets in Fig.~\ref{fig:Wild_EPJB2010_IVCs2050} show the magnetic
field dependence of the critical current of the SIFS Josephson junctions
measured at $500\,$mK. The dependencies are close to a Fraunhofer diffraction
pattern
\begin{eqnarray}
I_{\rm c}(H) & = & I_{\rm c}(0) \left| \frac{ \sin (\pi \Phi / \Phi_0)}{\pi \Phi / \Phi_0} \right|
\label{eq:IcH}
\end{eqnarray}
expected for an ideal short Josephson junction with a spatially uniform $J_{\rm
c}$. This demonstrates that
our SIFS junctions have good uniformity of $J_{\rm c}$ across the junction
area, that is, a spatially homogeneous tunneling barrier and ferromagnetic
interlayer. In particular, there are no short circuits at the junction edges or
the surrounding SiO$_2$ wiring insulation. We note that direct information on
junction inhomogeneities on smaller length scales can be obtained by Low
Temperature Scanning Electron
Microscopy~\cite{GrossRPP1994,BoschPRL1985,FischerScience1994,GerdemannJAP1994}.

We note that the total flux $\Phi$ threading the junction is composed of the
flux $\Phi_{\rm H}$ due to the applied external magnetic field and the flux
$\Phi_{\rm M}$ due to the magnetization of the ferromagnetic layer. The two
components are given by
\begin{eqnarray}
\Phi_{\rm H} & = & \widetilde{t}_{\rm B}^{\rm j} L^{\rm j} \; \mu_0 H \;\; ,
\label{eq:phiH_a}
\\
\Phi_{\rm M} & = & d_{\rm F} L^{\rm j} \; \mu_0 M \;\; .
\label{eq:phiH_b}
\end{eqnarray}
Here, $L^{\rm j} $ is the lateral dimension of the rectangular junction
perpendicular to the field direction, $d_{\rm F}$ the thickness of the F
interlayer, and $\widetilde{t}_{\rm B}^{\rm j}$ the effective magnetic
thickness of the junction given by eq.(\ref{eq:effectivethickness2}). Since the
applied magnetic field $H_{\Phi_0} = \Phi_0/\mu_0L^{j}\widetilde{t}_{\rm B}^{\rm
j}$ required for the generation of a single flux quantum in the junction is
less than about 1\,mT for $10\,\mu\mathrm{m} \le L^{\rm j} \le
50\,\mu\mathrm{m}$, the typical field range used in the measurement of the
$I_{\rm c}(\Phi/\Phi_0)$ curves of Fig.~\ref{fig:Wild_EPJB2010_IVCs2050} is
restricted to less than about 10\,mT. For such small in-plane magnetic fields
the $M(H)$ curve of the ferromagnetic PdNi layer (cf.
Fig.~\ref{fig:Wild_EPJB2010_SQUID}) can be well approximated by a linear
dependence $M \simeq \mu H$ with $\mu = 1.8$. With this approximation the total
magnetic flux threading the junction can be expressed as
\begin{eqnarray}
\Phi & = &  \mu_0 H \; L^{j} (\widetilde{t}^{\rm j}_{\rm B} -d_{\rm F}) + \mu_0 \mu H \; L^{\rm j} d_{\rm F}
\nonumber \\
 & = & \mu_0 H \; L^{\rm j} \widetilde{t}_{\rm B}^{\rm j} \; \left( 1 + (\mu -1) \, \frac{d_{\rm F}}{\widetilde{t}_{\rm B}^{\rm j}} \right)
 \;\; .
\label{eq:phitotal}
\end{eqnarray}
Here, $M$ is the magnetization component parallel to the applied magnetic
field. If there is a significant in-plane magnetic anisotropy of the
ferromagnetic material, this component may be much smaller than the absolute
value of the magnetization. Furthermore, there may be a complicated domain
structure. In this case, $M$ is the average magnetization parallel to the field
direction.

We can use (\ref{eq:phitotal}) to estimate the magnitude of the additional flux
$\Phi_{\rm M}$ due to the F layer. For the junctions of
Fig.~\ref{fig:Wild_EPJB2010_IVCs2050}, $d_{\rm F} = 8.4$\,nm and hence $d_{\rm
F}/\widetilde{t}_{\rm B}^{\rm j} \simeq 0.05$. Therefore, for $\mu = 1.8$ the
second summand in the brackets of (\ref{eq:phitotal}) amounts to only about
0.04. That is, compared to a junction without F interlayer the total flux is
enhanced only by about 4\%. Due to the uncertainties in $\lambda_{\rm L}$ and
the geometrical dimensions of the junctions, this small effect is difficult to
prove. Furthermore, since the virgin part of the hard axis $M(H)$ curve of the
PdNi layer is about linear at small fields with negligible hysteresis, the
$I_{\rm c}(\Phi/\Phi_0)$ curves are expected to show negligible hysteresis on
sweeping back and forth the applied magnetic field. This is in agreement with
our data and measurements on SIFS junctions with ferromagnetic CuNi
interlayers~\cite{Weides2008}. We note, however, that $\Phi_{\rm M}$ strongly
depends on magnetic history. After having increased the applied field to large
values saturating the F layer, there is a significant remanent magnetization on
reducing the field to zero again. This remanence causes a large $\Phi_{\rm M}$
which may exceed $100\Phi_0$. Unfortunately, applying high magnetic fields also
results in the trapping of Abrikosov vortices in the Nb junction electrodes,
making the interpretation of the measured $I_{\rm c}(\Phi/\Phi_0)$ dependencies
difficult.

\section{Fiske Resonances}
\label{sec:Fiske}

\begin{figure}[tb]
  \centering
 \includegraphics[width=.95\columnwidth]{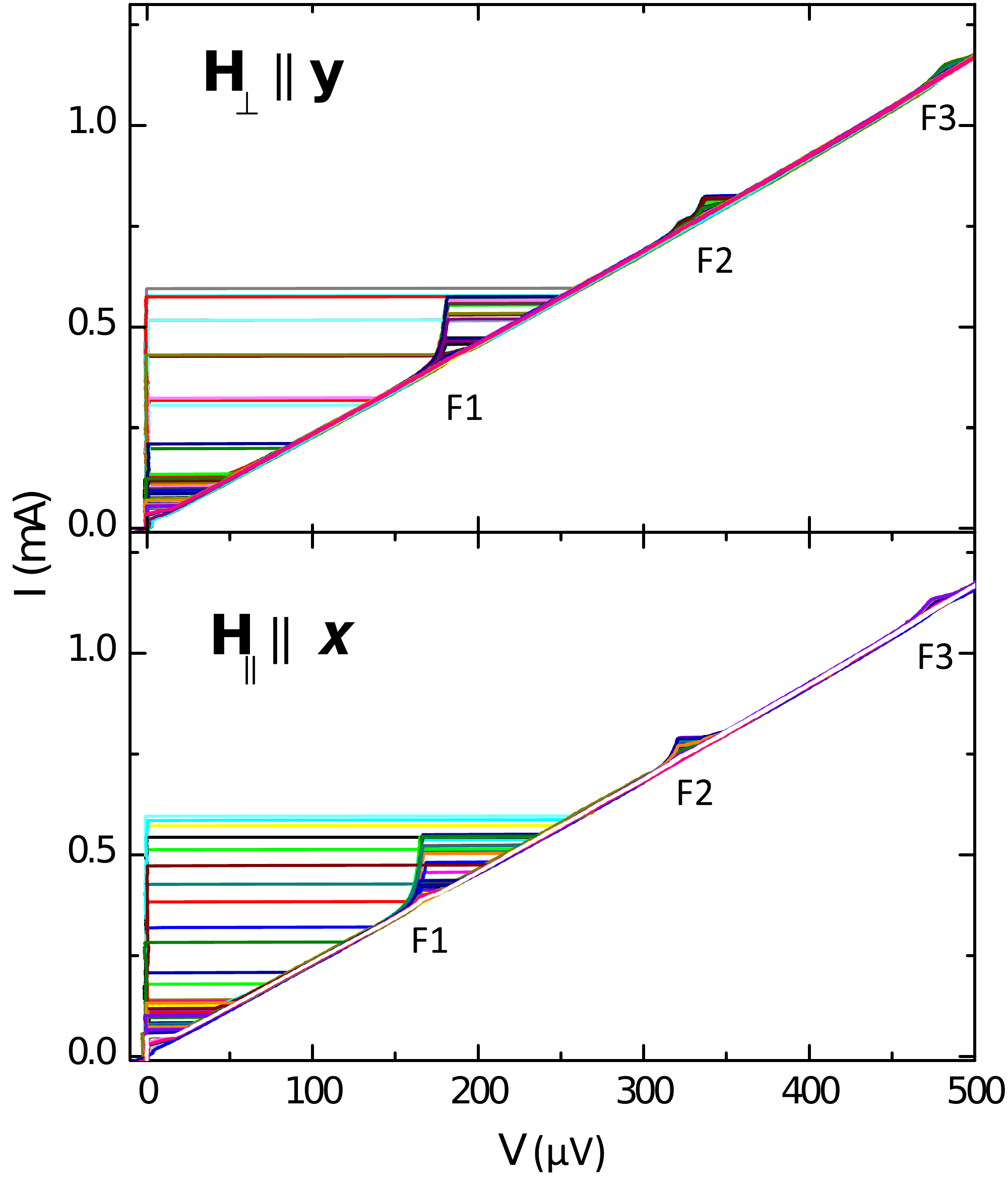}
 \caption{
Current-voltage characteristics (IVCs) of a SIFS Josephson junction with
$A=50\times 50\,\mu\mathrm{m}^2$ and $d_{\rm F}=8.4$\,nm measured at 500\,mK for different magnetic
fields applied parallel (bottom) and perpendicular (top) to the bottom
electrode. The first three Fiske resonances are clearly seen and labeled F1, F2, and F3.}
 \label{fig:Wild_EPJB2010_IVCs50}
\end{figure}

\begin{figure}[tb]
  \centering
 \includegraphics[width=.95\columnwidth]{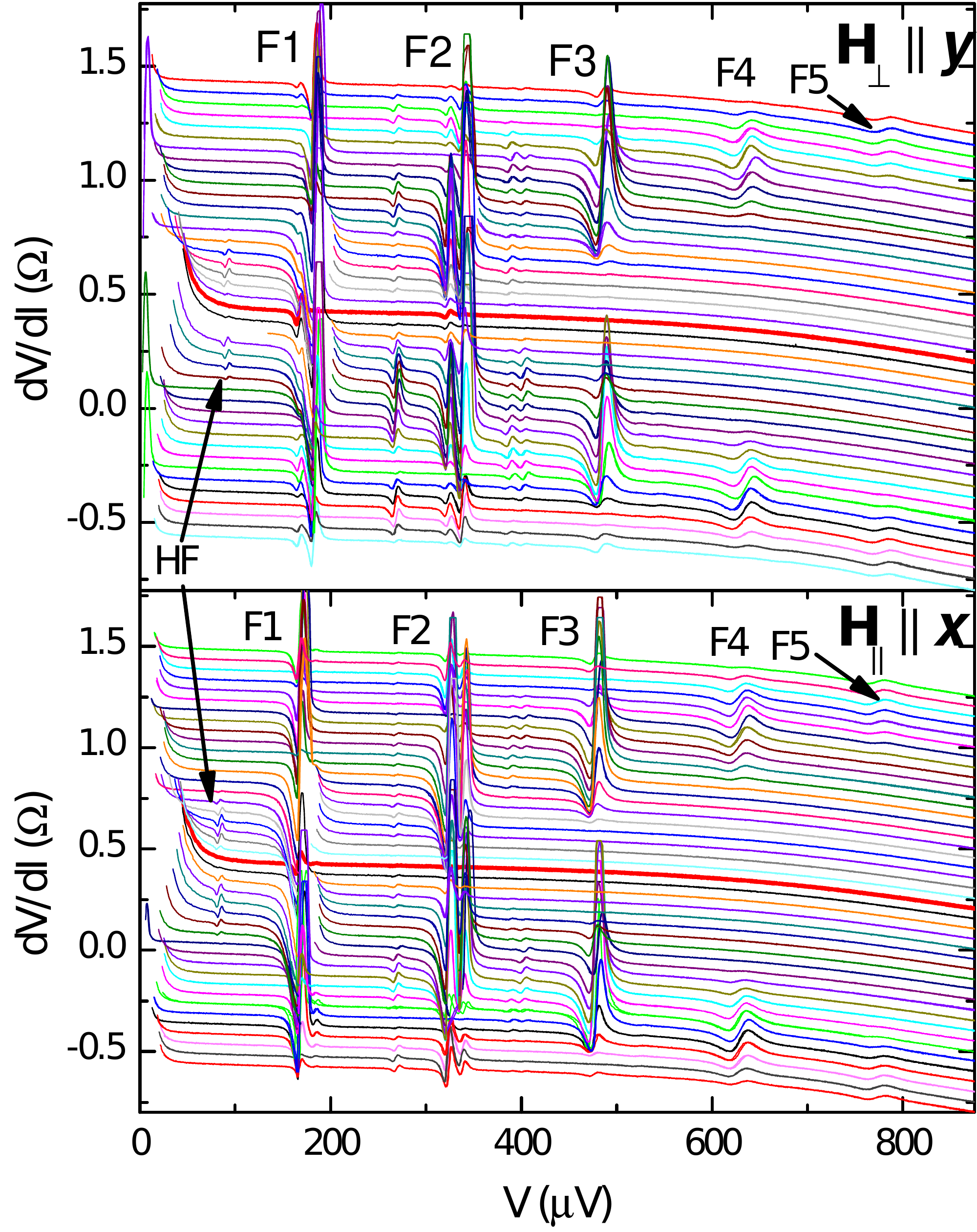}
 \caption{
Differential resistance $dV/dI$ vs.\ applied voltage $V$ measured with a
lock-in technique at 500\,mK for different magnetic fields applied parallel
(bottom) and perpendicular (top) to the bottom electrode. The data is taken
simultaneously to those shown in Fig.~\ref{fig:Wild_EPJB2010_IVCs50}. Fiske
resonances are observed up to almost $800\,\mu$V corresponding to a frequency of
about 400\,GHz. The origin of the resonance labeled HF at half the first Fiske
voltage is discussed in the text. The thick red lines mark the measurements
taken at zero applied magnetic field. These curves are not shifted vertically,
whereas all other curves are subsequently shifted by $0.1\,\Omega$ for clarity.}
 \label{fig:Wild_EPJB2010_URs50}
\end{figure}

According to the Josephson equations, the supercurrent across a Josephson
junction oscillates at a constant frequency $\omega_{\rm J} = 2\pi V/\Phi_0$
when a constant voltage $V$ is applied across the junction~\cite{Josephson}. On
the other hand, the junction geometry forms a transmission line resonator of
length $L^{\rm j}$ (cf. Fig.~\ref{fig:Overlap}) with resonance frequencies
$\omega_n/2\pi = n\,\frac{\bar c}{2L^{\rm j}}$. Here, $n = 1,2,3, \ldots$ and
$\bar c$ is the Swihart velocity~\cite{Swihart61}. At an applied junction
voltage $V_n = \omega_n \Phi_0/2\pi = n \, \frac{\Phi_0 \bar
  c}{2L^{\rm j}}$, the oscillation frequency of the Josephson current matches
the $n$-th harmonic of the junction cavity mode, potentially resulting in the
excitation of the cavity modes. The excitation of the cavity resonances is most
effective, if the spatial period of the Josephson current distribution along the
junction is about matching the spatial period of the $n$-th resonant
electromagnetic mode of the junction. Since for short junctions the Josephson
current density is spatially uniform at zero magnetic field, there is no
excitation of the resonant modes. However, a spatial modulation can be easily
achieved by applying a magnetic field. The resonance is a highly nonlinear
process which results in self-induced current steps at $V_n$ in the IVCs
denoted as Fiske steps~\cite{PhysRevFiske,RMPFiske,Barone}. These steps are
clearly visible in the IVCs shown in Fig.~\ref{fig:Wild_EPJB2010_IVCs50} and
their derivatives $dV/dI$ plotted in Fig.~\ref{fig:Wild_EPJB2010_URs50}. The
data were obtained for a junction with area $A=50\times 50\,\mu\mathrm{m}^2$
recorded at different magnetic fields applied parallel ($x$-direction) and
perpendicular ($y$-direction) to the bottom electrode. The $dV/dI$ vs. $V$
curves were measured by a lock-in technique. In the case of negligible damping
the resonances are very sharp at almost constant voltages. However, in reality
the resonances are damped due to different loss mechanisms. For planar type SIFS
junctions as studied in our work the junction damping due to radiation losses is
small due to the large electromagnetic impedance mismatch at the junction
boundaries. Then, the quality factors $Q_n$ of the Fiske resonances are
given mainly by the internal losses, most likely due to quasiparticle tunneling
and the finite surface resistance. In general, the detailed analysis of the
Fiske resonances can provide information on the damping mechanisms in Josephson
junctions at very high frequencies.

When analyzing the voltage position of the Fiske resonances in detail we have
to take into account the idle region next to the junction area. As discussed
above, this idle region results in an increase or decrease of the phase
velocity $v_{\rm ph}$ compared to the Swihart velocity $\bar{c}$ of an ideal
junction without any idle region, depending on whether the idle region is
lateral or longitudinal, respectively. In turn, this causes an increase or
decrease of the characteristic voltages $V_n = n \, \frac{\Phi_0 v_{\rm
ph}}{2 L^{\rm j}}$. For $H \| y$, the resonant modes are extending in
$x$-direction, that is, parallel to the idle region (lateral mode). In this
case the effect on the phase velocity is described by eq(\ref{eq:Swihart10})
and we expect a slight increase of $v_{\rm ph}$ and $V_n$. Analogously,
for $H\| x$ the resonant modes are extending in $y$-direction, that is,
perpendicular to the idle region (longitudinal mode). In this case, the effect
on the phase velocity is described by eq(\ref{eq:Swihart11}) and we expect a
slight decrease of $v_{\rm ph}$ and $V_n$. This is in good qualitative
agreement with our experimental observation. For $H \| x$, resonant modes are
found at $167\,\mu$V, $322\,\mu$V, $475\,\mu$V, $630\,\mu$V and $772\,\mu$V,
whereas for $H \| y$ the resonant modes appear at slightly larger voltages
$181\,\mu$V, $336\,\mu$V, $483\,\mu$V, $636\,\mu$V and $780\,\mu$V. We note,
that the correction factors estimated from eqs.(\ref{eq:Swihart10}) and
(\ref{eq:Swihart11}) indicated an even bigger difference between the lateral and
longitudinal mode. However, the quantitative evaluation of this difference
depends on the details of the junction geometry and properties of the involved
materials (e.g. dielectric constants) and therefore needs a more elaborate
effort~\cite{Lee91,IEEEApplSup2Lee}. Qualitatively we can say that the effect
of the idle region is reduced on going to higher frequencies (smaller
wavelengths) due to a stronger confinement of the modes in the junction area.
That is, the difference of the resonance voltages of the lateral and
longitudinal mode is decreasing on going to higher harmonics. This nicely
agrees with our observation and reports in literature~\cite{Lee91}.

The Fiske steps $\Delta I_n(V, \Phi)$ are obtained from the measured IVCs
(cf. Fig.~\ref{fig:Wild_EPJB2010_IVCs50}) by subtracting an ohmic background
which is determined by the about constant subgap resistance. Equivalently,
$\Delta I_n(V, \Phi)$ can be obtained by integration of the $dV/dI$ curves
shown in Fig.~\ref{fig:Wild_EPJB2010_URs50}, again after substraction of the
subgap resistance. The latter method yields better results for the higher
modes. A detailed quantitative description of the voltage position and the shape
of the Fiske resonances was given by Kulik for standard SIS Josephson tunnel
junctions~\cite{JETPKulik}. In this section we analyze the Fiske steps $\Delta
I_n(V,\Phi)$ in the IVCs of our SIFS Josephson junctions using an
extension of Kulik's
theory~\cite{Barone,Kulik1967,Paterno1978,PRBGou}. According to this theory, the
Fiske steps can be expressed as~\cite{Barone,Kulik1967}
\begin{eqnarray}
\Delta I_n(V,\phi) & = & \frac{I_{\rm c}}{4\pi^2n^2} \left(\frac{L^{\rm j}}{\lambda_{\rm J}} \right)^2
\left(\frac{V_n}{V} \right)^2
\nonumber \\
& &
\sum\limits_{n=0}^\infty \frac{1/Q_n}{\left[ 1- \left(\frac{V_n}{V}
\right)^2 \right]^2 + \left(\frac{1}{Q_n}\right)^2 } \; F_n^2
(\phi)
 \; .
\label{eq:Fiske1}
\end{eqnarray}
Here, $Q_n$ is the quality factor of the $n$-th resonant mode and $V_n$
its voltage position. The function $F_n(\phi)$ describes the flux
dependence of the resonant mode given by
\begin{eqnarray}
F_n(\phi) & = & \frac{2\phi \cos (\pi\phi)}{\pi \left[ \phi^2 - (n/2)^2 \right] } \hspace*{5mm} \textrm{for} \;\; n = 1,3,5,\ldots
\label{eq:Fiske3a}
\\
F_n(\phi) & = & \frac{2\phi  \sin (\pi\phi)}{\pi \left[ \phi^2 - (n/2)^2
\right] } \hspace*{5mm}  \mathrm{for} \;\; n = 2,4,6,\ldots
\label{eq:Fiske3b}
\end{eqnarray}
with $\phi = \Phi/\Phi_0$. The flux dependence of the maximum height of
the Fiske steps is given by~\cite{Barone,Kulik1967}
\begin{eqnarray}
\Delta I_n^{\rm max}(\phi^{\rm max}) & = &
I_c \, \left(\frac{L^{\rm j}}{\lambda_{\rm J}}\right)^2 \; \frac{Q_n}{4\pi^2 n^2} \; F_n^2 (\phi_n^{\rm max})
 \; .
\label{eq:Fiske2}
\end{eqnarray}
Fitting the current steps $\Delta I_n(V)$ measured at fixed applied flux
to (\ref{eq:Fiske1}) allows us to determine the voltage positions $V_{n,
\|}$ ($V_{n, \perp}$) and the maximum step heights $\Delta I_{n,
\|}^{\rm max}$ ($\Delta I_{n, \perp}^{\rm max}$) for the magnetic field
applied parallel (perpendicular) to the base electrode. For low $n$, the
quality factors determined by eq.(\ref{eq:Fiske2}) are not within the validity
of Kulik's theory applicable only for low $Q$. Therefore, the the quality
factors $Q_{n, \|}$ ($Q_{n,\perp}$) listed in Table~\ref{tab:FiskeQ}
have been calculated by an extended theory more appropriate for the high-$Q$
regime~\cite{Barone}.

\begin{table*}[tb]
\caption{ Voltage position $V_{n, \|}$ ($V_{n, \perp}$) and step height
$\Delta I_{n, \|}^{\rm max}$ ($\Delta I_{n, \perp}^{\rm max}$) of the
$n$-th Fiske steps obtained for the magnetic field applied parallel
(perpendicular) to the bottom electrode. The data was derived from the IVCs of
a $50\times 50\,\mu\mathrm{m}^2$ junction measured at 500\,mK. Also listed are
the quality factors $Q_{n, \|}$ and $Q_{n, \perp}$ of the $n$-th Fiske
resonance, which are obtained by fitting the measured $\Delta I_{n,
\|}(V,\Phi)$ and $\Delta I_{n, \perp}(V,\Phi)$ dependencies by Kulik's
theory~\cite{JETPKulik,Kulik1967}.}
\begin{center}
  \begin{tabular}{lcccccc}
   \hline\hline\noalign{\smallskip}
    number $n$ of Fiske step  \hspace*{40mm}     & 1/2 & 1 & 2 & 3 & 4 & 5 \\
    \noalign{\smallskip}\hline\hline\noalign{\smallskip}
     $H \|$ bottom electrode ($x$-direction):          & & & & &  &  \\
    $\Delta I_{n, \|}^{\rm max}$ ($\mu$A)    & & 140 & 48 & 27 & 13 & 6 \\
    $Q_{n \|}$                & & 22 & 27 & 35 & 30 & 22\\
    $V_{n, \|}$ ($\mu$V)    & 83.5 &167 & 322 & 475 & 630 & 772\\
    \noalign{\smallskip}\hline\noalign{\smallskip}
    $H \perp$ bottom electrode ($y$-direction):        & & & & &  &  \\
    $\Delta I_{n, \perp}^{\rm max}$ ($\mu$A) & & 130 & 46 & 24 & 13 & 5 \\
    $Q_{n, \perp}$                 & & 19 & 26 & 31 & 30 & 18 \\
    $V_{n, \perp}$ ($\mu$V)   & 90.5 & 181 & 336 & 483 & 636 & 780 \\
%    \hline
%    $Q_q=Q \cdot n \cdot \frac{V_1}{\Phi_0 \cdot \omega_p}$ $0^\circ$  && 37 &
%    74 & 111 & 148 & 185 \\
%    12*n*160e-6/phi0/11.8GHz
%    $Q_r 0^\circ$ & & 300 & 155 & 176 & 93 & 53 \\
     \noalign{\smallskip}\hline
  \end{tabular}
\end{center}
    \label{tab:FiskeQ}
\end{table*}

\begin{figure}[tb]
  \centering
 \includegraphics[width=.95\columnwidth]{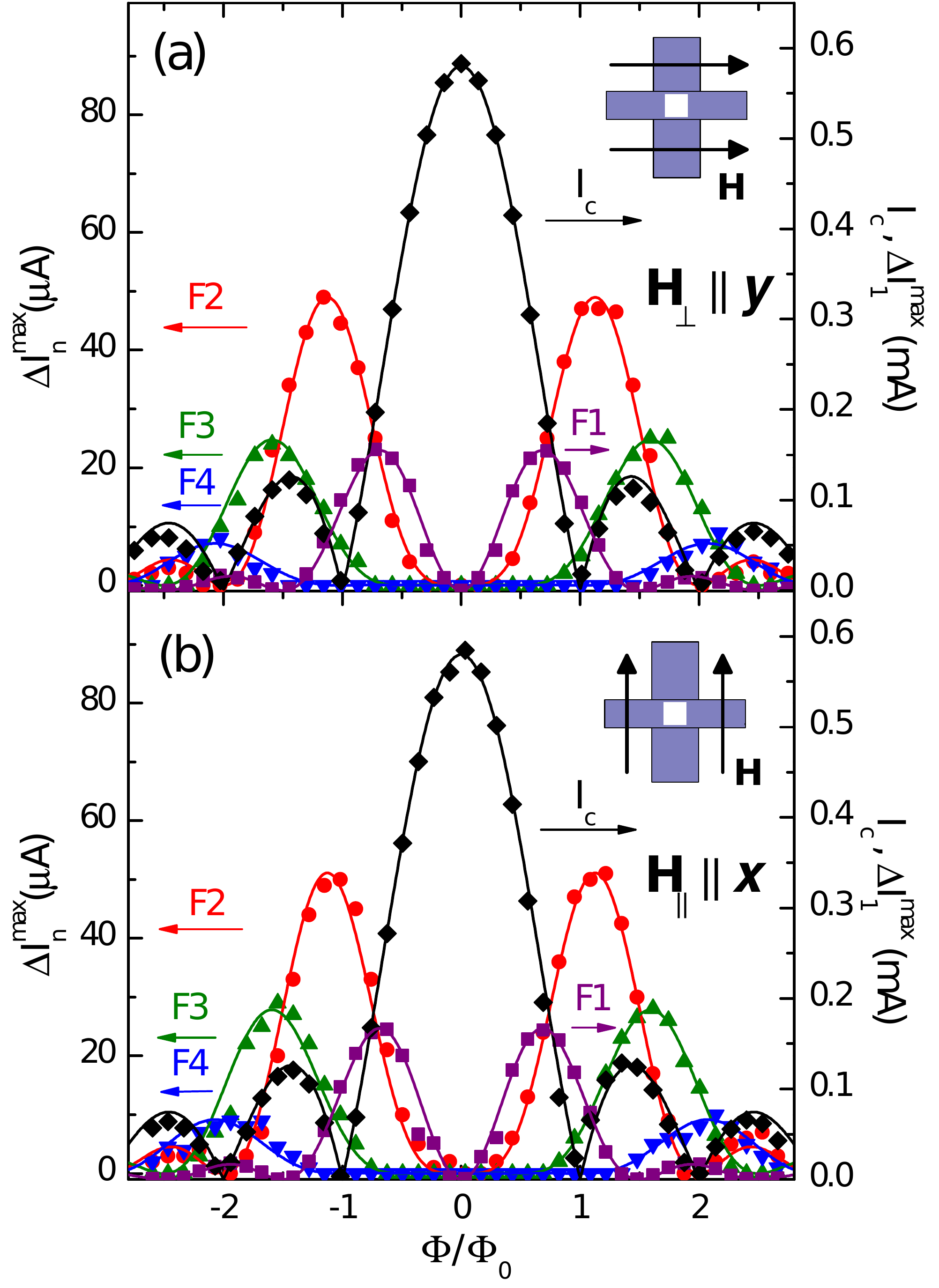}
 \caption{
Dependence of the critical current $I_{\rm c}$ and the height $\Delta I_{\rm n}^{\rm max}$ of the
Fiske steps on the magnetic flux generated by a magnetic field applied
perpendicular (a) and parallel (b) to the bottom electrode. The data is
obtained for a SIFS Josephson junction with $A=50\times 50\,\mu\mathrm{m}^2$ at
500\,mK. For clarity, the magnetic field dependence of $I_{\rm c}$ and the
height of the first Fiske step (F1) are linked to the right axis. The data is
fitted to a Fraunhofer pattern and Kulik's theory~\cite{JETPKulik,Kulik1967}.}
 \label{fig:Wild_EPJB2010_FiskeIB}
\end{figure}

As discussed above, for short Josephson junctions the excitation of the cavity
resonances is most effective, if the spatial period of the Josephson current
distribution along the junction is about matching the spatial period of the
$n$-th resonant electromagnetic mode of the junction. Therefore, the height of
the Fiske steps strongly depends on the applied magnetic field and has a
pronounced maximum at a particular field value. This is shown in
Fig.~\ref{fig:Wild_EPJB2010_FiskeIB}, where we have plotted the height $\Delta
I_n^{\rm max}$ of the $n$-th Fiske step together with the critical
current $I_{\rm c}$ of a $50\times 50\,\mu\mathrm{m}^2$ junction versus the
applied magnetic flux. For large $n$, the Fiske step has a maximum height at
$\Phi \simeq n\Phi_0/2$ where the Josephson current shows about the same
spatial modulation along the junction as the cavity mode.

The voltage positions $V_n$ of the Fiske resonances allow us to derive
several interesting junction parameters such as the Swihart velocity, the
specific capacitance, or the junction plasma frequency. To avoid any
ambiguities related to the overlap effect discussed above, we use the Fiske
voltages found for $H \| x$ to derive these parameters. Obviously, the values
$V_n = n \, \frac{\Phi_0 \bar c}{2 L^{\rm j}}$ directly give the Swihart
velocity $\bar c$. From the position $V_1=165\,\mu$V of the first Fiske step we
obtain $\bar c$ to $0.027 \cdot c$, where $c$ is the speed of light in vacuum.
We note that the derived Swihart velocity may slightly vary with increasing
step number, e.g. due to a nonvanishing dispersion of the dielectric constant
of the barrier material~\cite{IEEEApplSup2Lee,PRBHermon}. Using the Josephson
penetration depth $\lambda_{\rm J} \simeq 70\,\mu$m determined in
section~\ref{sec:properties}, the plasma frequency $\omega_{\rm p} / 2\pi =
\bar c / \lambda_{\rm J}$ is obtained to 17.8\,GHz. Since $\omega_{\rm p} =
\sqrt{2eJ_{\rm c}/\hbar C_{\rm s}}$, we can derive the specific capacitance
$C_{\rm s}$ of the junction to $C_{\rm s}=54\,\mathrm{fF}/\mu\mathrm{m}^2$.
This value is about two times larger than typical values reported in literature
for Nb/AlO$_x$/Nb Josephson junctions~\cite{APLZant}. With these values the
quality factor $Q_{\rm Swihart}=\omega_{\rm p} R_{\rm p} A C_{\rm s}$ can be
estimated, which directly follows from the Swihart velocity derived from the
position of the first Fiske resonance. Here, $R_{\rm p}$ is the resistance of
the junction measured at a voltage  $V_{\rm p}=\frac{\omega_{\rm p}
\Phi_0}{2\pi} \simeq 35\,\mu$V corresponding to the plasma frequency. $R_{\rm
p}$ agrees well with the subgap resistance $R_{\rm sg}= 0.44\,\Omega$. Using
this value we obtain the quality factor $Q_{\rm Swihart}=6.6$, which
corresponds to the quality factor $Q_{\rm IVC}$ determined from the retrapping
current of the IVC of the same junction using the RCSJ-model. The fact that
$Q_{\rm Swihart}$ is slightly larger than the value $Q_{\rm IVC} =5.7$ is not
astonishing keeping in mind effects of premature switching/retrapping due to
noise in the IVC measurement. For the $20\times 20\,\mu\mathrm{m}^2$ junction
we observed Fiske resonances at $V_1=427\,\mu$V and $V_2= 790\,\mu$V (not
shown). Taking into account a slight increase of the first Fiske voltage due to
the overlap effect and a decrease of the second due to dispersion effects, we
can estimate a Fiske step distance of about $415\,\mu$V. This value leads to
$C_{\rm s} = 53\,\mathrm{fF}/\mu\mathrm{m}^2$ in agreement with the value
obtained for the large junction. The derived plasma frequency is $\omega_{\rm
p} /2\pi=21.7\,$GHz and the quality factor is $Q_{\rm Swihart} =6.2$.

We next use the quality factors derived from the Fiske steps to analyze damping
effects. In the theoretical description of the Fiske resonances a finite
damping is assumed. However, the origin of this damping is usually not
specified. It turns out that the experimental values of $Q_n$ derived
from the Fiske resonances can be both larger and smaller than the quality
factor $Q_{\rm IVC}$ derived from the resistively and capacitively shunted
junction (RCSJ) model. The reason is that on the one hand the RCSJ model
overestimates the losses due to quasiparticle tunneling, since a voltage
independent resistance is assumed in this model. On the other hand, the RCSJ
model does not take into account other loss mechanisms, e.g. due to a finite
surface resistance. Assuming that there are various loss mechanisms the total
quality factor of the Fiske resonances can be written as
\begin{eqnarray}
 Q_n & = & \left( \frac{1}{Q_{n,\rm qp}} +
 \underbrace{\frac{1}{Q_{n,\rm rad}} +  \frac{1}{Q_{n,\epsilon}}
 + \frac{1}{Q_{n,R_{\rm s}}} + \frac{1}{Q_{n,L}}}_{1/\widetilde{Q}_n} \right)^{-1} \; .
\label{eq:Qn}
\end{eqnarray}
Here, the different contributions represent losses due to quasi-particle
tunneling ($Q_{n,\rm qp}$) and a finite surface resistance ($Q_{n,R_{\rm s}}$) as
well as radiation losses ($Q_{n,\rm rad}$), and dielectric losses ($Q_{n,\rm
  \epsilon}$)~\cite{PRBSmith}. Furthermore, variations $\Delta L$ in the
junction length $L$ lead to a broadening of the resonances which can be
expressed by a frequency independent quality factor $Q_{n,L} = L^{\rm
  j}/\Delta L^{\rm j}$.

\begin{figure}[tb]
  \centering
 \includegraphics[width=.95\columnwidth]{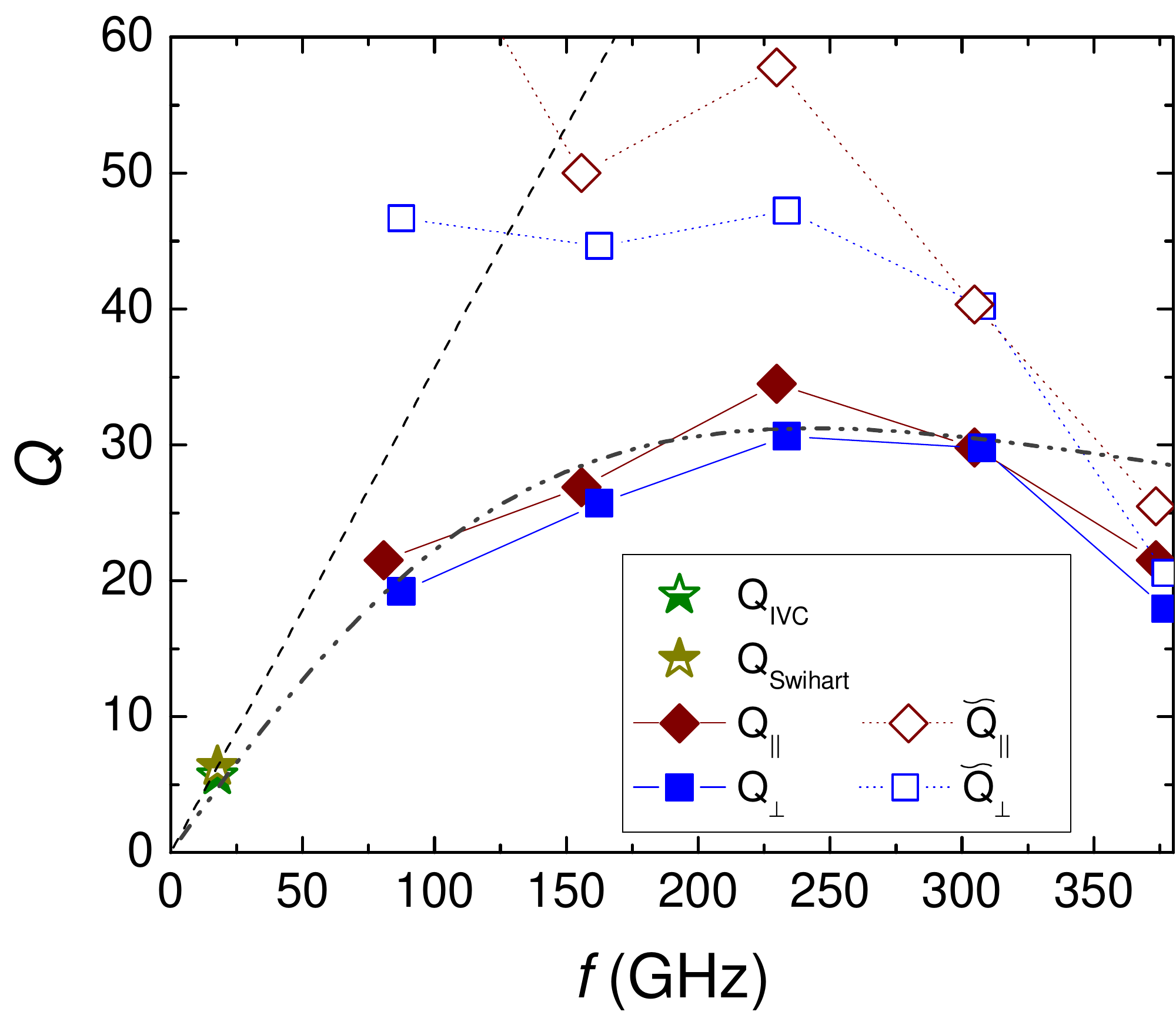}
 \caption{
Quality factors $Q_{\rm IVC}$ and $Q_{\rm Swihart}$ derived from the IVCs and
the Swihart velocity by assuming only quasiparticle damping. The broken
straight line gives an extrapolation of these values to higher frequencies.
Also shown are the quality factors $Q_{\|}$ and $Q_{\perp}$ derived from
the Fiske resonances (full symbols). The open symbols represent the quality
factors $\widetilde{Q}_{\|}$ and $\widetilde{Q}_{\perp}$ derived from
the Fiske resonances after subtraction of the extrapolated quasiparticle
damping. The dash-dotted line is obtained by fitting the data using (\ref{eq:Qnfit}).}
 \label{fig:Wild_EPJB2010_Q}
\end{figure}

If quasiparticle tunneling is the only damping mechanism we would expect
$Q_{n,\rm qp} = \omega_n RC$. Here, $R$ is the junction normal resistance
at the voltage $V_n = \hbar\omega_n/2e$, which is given by the
about constant subgap resistance $R_{\rm sg}$ in the relevant regime.
Therefore, $Q_{n,\rm qp}$ is expected to increase about linearly with
increasing resonant mode frequency. Above we have determined
$Q_{\rm Swihart} = \omega_{\rm p} RC= 6.6$. Extrapolating the value to larger
frequency we would expect $Q_{n,\rm qp} = 0.37\cdot(\omega/2\pi)\,{\rm GHz}^{-1} $. This dependence
is shown in Fig.~\ref{fig:Wild_EPJB2010_Q} by the broken straight line.
Obviously, the quality factors determined from the Fiske resonances do not
follow this line. From this we can conclude that there are additional damping
mechanisms beside quasiparticle tunneling. In order to get some insight into
the frequency dependence of these additional mechanisms, we have used
(\ref{eq:Qn}) to determine the part in the quality factor due to the additional
damping mechanisms by subtracting the quasiparticle damping. The resulting
values $\widetilde{Q}_n$ are shown as open symbols in
Fig.~\ref{fig:Wild_EPJB2010_Q}. It is evident that $\widetilde{Q}_n$
decreases significantly with increasing frequency.

As discussed above there are several possible mechanisms leading to additional
damping in Josephson junctions. However, radiation and dielectric losses
usually can be neglected due to the large impedance mismatch at the junction
boundaries and the small volume of the dielectric, respectively. Furthermore,
in our fabrication process, $L^{\rm j}/\Delta L^{\rm j} \simeq 100$ for $L^{\rm
j}=50\,\mu\mathrm{m}$ resulting in  $Q_{n,L} \simeq 100$. That is, the
geometric inhomogeneities set an upper limit of the quality factor of the
junctions, which is above the measured values. The remaining mechanism is
damping due to the finite surface resistance of the junction electrodes.
Expressing the complex surface impedance of a superconductor as $Z_{\rm s} =
R_{\rm s} + \imath X_{\rm s}$, the contribution of the junction electrodes to
the quality factor is given by the ratio~\cite{PRBBroom}
\begin{eqnarray}
Q_{n,R_{\rm s}} & = &  \frac{X_{\rm s}}{R_{\rm s}} \; .
\label{eq:QRs1}
\end{eqnarray}
For a rough estimate of $Q_{n,R_{\rm s}}$ we can use a simple two-fluid model.
Expressing the conductivity of the superconductor as the sum $\sigma =
\sigma_{\rm s} + \sigma_{\rm n}$ of the superconducting and normal conducting
carriers and using the simple relations \cite{Tinkham}
\begin{eqnarray}
\sigma_{\rm s} & = &  \frac{1}{\imath \omega \mu_0 \lambda_{\rm L}^2}
\label{eq:sigma1a}
\\
\sigma_{\rm n} & = &  \frac{n_{\rm n} e^2}{m_{\rm n}} \; \frac{\tau}{1+\imath \omega \tau}
\label{eq:sigma1b}
\end{eqnarray}
we can derive the following expression for the real and imaginary part of the
complex surface impedance:
\begin{eqnarray}
R_{\rm s} & = &  \frac{1}{2} \omega^2 \mu_0^2 \lambda_{\rm L}^3 \sigma_{\rm N} \left(\frac{n_{\rm n}}{n}\right)
\label{eq:Zsreal}
\\
X_{\rm s} & = &  \omega \mu_0 \lambda_{\rm L} \; .
\label{eq:Zsimaginary}
\end{eqnarray}
Here, $n_{\rm n}/n$ is the temperature dependent fraction of normal electrons,
$m_{\rm n}$ the mass of the normal electrons, and $\sigma_{\rm N}$ is the
conductivity of the superconductor in the normal state where $n_{\rm n} = n$.
With these expressions we obtain
\begin{eqnarray}
Q_{n,R_{\rm s}} & = &  \frac{2}{\omega \mu_0\lambda_{\rm L}^2 \sigma_{\rm N} \left( n_{\rm n}/n \right) } \; .
\label{eq:QRs2}
\end{eqnarray}
We see that $Q_{n,R_{\rm s}}$ increases strongly with decreasing temperature due
to the freeze out of the normal electrons. However, due to the uncertainties in
$n_{\rm n}/n$, $\sigma_{\rm N}$ in the PdNi/Nb bilayer, and the simplicity of
our approach, eq.~(\ref{eq:QRs2}) certainly cannot be used to estimate the
absolute value of $Q_{n,R_{\rm s}}$. However, according to (\ref{eq:QRs2}) we
expect $Q_{n,R_{\rm s}} \propto 1/\omega$. This explains the observed decrease of
$\widetilde{Q}_n$ with increasing frequency shown in
Fig.~\ref{fig:Wild_EPJB2010_Q}.

With the quasiparticle tunneling, the finite surface resistance and geometric
inhomogeneities as the three main contributions to the measured quality factor,
we expect
\begin{eqnarray}
 \frac{1}{Q_n} & = & \frac{1}{Q_{n,\rm qp}^0 \; \omega}
 + \frac{\omega }{Q_{n,R_{\rm s}}^0} + \frac{1}{Q_{n,L}} \; .
\label{eq:Qnfit}
\end{eqnarray}
As shown in Fig.~\ref{fig:Wild_EPJB2010_Q}, this expression well fits the
measured data with $Q_{n,\rm qp}^0=0.046\times 10^{-9}\mathrm{s}$, $Q_{n,R_{\rm
    s}}^0 = 2500\times 10^9\mathrm{s}^{-1}$, and $Q_{n,L} = 100$. From this
we can learn that the quality factor of our SIFS junctions is limited by
quasiparticle tunneling at low frequencies and the finite surface resistance at
high frequencies. In the intermediate regime there may be an effect of geometric
inhomogeneities when going to small area junctions.

The quality factors measured for our SIFS Josephson junctions are slightly
lower than the values reported for Nb/AlO$_x$/Nb tunnel
junctions~\cite{IEEEGijsbertsen}. This is not astonishing because there is the
additional F layer in our SIFS junctions. First this layer reduces the $I_{\rm
c}R_{\rm n}$ product of the SIFS junction compared to a SIS junction and
thereby increases the effect of quasiparticle damping characterized by $Q_{\rm
IVC} \simeq Q_{\rm Swihart} = \omega_{\rm p} R_{\rm n}C = \sqrt{2eI_{\rm
c}R_{\rm n}^2C/\hbar}$. Furthermore, the F layer results in an increased
surface resistance since the PdNi/Nb bilayer has increased $\sigma_{\rm N}$ and
an increased fraction $n_{\rm n}/n$ of the normal electrons. The increased
$n_{\rm n}/n$ value is a result of the inverse proximity effect, increasing the
quasi-particle density in the superconducting electrodes~\cite{Bergeret2005a}.
Finally, effects originating from magnetic impurity scattering on Ni atoms
diffused into the niobium top layer may play a role~\cite{Nam:1967}.

Comparing SIFS to SFS junctions, it is immediately evident that the quality
factors of SFS junctions will be very small. Due to the much smaller normal
resistance and vanishing capacitance, SFS junctions are overdamped ($Q<1$). The
larger quality factors of the SIFS junctions are obtained by the additional
tunneling barrier, which causes large $R_{\rm n}$ and $C$. Increasing the
thickness $t^{\rm j}$ of the tunneling barrier is expected to result in an
exponential increase of $R_{\rm n}$, while $I_{\rm c}R_{\rm n}$ should stay
constant and $C$ should decrease as $1/t^{\rm j}$. Therefore, increasing
$t^{\rm j}$ in principle can be used to further increase the quality factors of
SIFS junctions. However, this is obtained at the cost of lower $J_{\rm c}$ and
$\omega_{\rm p}$ what may be a problem for some applications. The quality
factors between about 5 and 30 in the frequency regime between about 10 and
400\,GHz achieved in our experiments are already sufficient for applications in
quantum information circuits or for studies of macroscopic quantum tunneling.

We conclude the discussion of Fiske resonances by paying attention to the small
resonances labeled HF in Fig.~\ref{fig:Wild_EPJB2010_URs50}, which occur at
exactly half of the voltage of the first Fiske step. This observation reminds
us of the appearance of half-integer Shapiro steps at the 0-$\pi$-transition of
SFS junctions due to a second harmonic component in the current-phase
relation~\cite{PRLSellier}. However, since the F layer thickness of the SIFS
junctions studied in our work is about twice the thickness of the
0-$\pi$-transition, this scenario would require a significant F layer thickness
variation of a few nanometers~\cite{goldobin2Phi}. This is in contradiction to
the small rms roughness of our F layers. Furthermore, the presence of a
double-sinusoidal current-phase relation as the origin of the HF resonance is
in contradiction with the measured magnetic flux dependence of its height.
Clearly the measured oscillation period of the HF resonance cannot be mapped on
twice the period of the first Fiske step~\cite{Biedermann1997}. Nonequilibrium
effects may be a possible explanation for the observed HF
resonance~\cite{Argaman1999}. However, further experiments are required to
clarify this point.

\section{Conclusion}
\label{sec:Conclusion}

We have fabricated Nb/AlO$_x$/$\rm Pd_{0.82}Ni_{0.18}$/Nb (SIFS) Josephson
junctions with controllable and reproducible properties using a self-align
process. High current densities up to more than $30$\,A/cm$^2$ and $R_{\rm
n}\cdot A$ values above $130\,\Omega\mu\mathrm{m}^2$ have been achieved. The
$I_{\rm c}(\Phi /\Phi_0)$ dependencies are close to an ideal Fraunhofer
diffraction pattern demonstrating the good spatial homogeneity of the
junctions. The transition from $0$- to $\pi$-coupled junctions was observed for
a thickness $d_{\rm F} \simeq 6$\,nm of the $\rm Pd_{0.82}Ni_{0.18}$ layer. The
$\rm Pd_{0.82}Ni_{0.18}$ layers show an out-of-plane anisotropy. They have a
Curie temperature of 150\,K, an exchange energy $E_{\rm ex} \simeq 20$\,meV,
and a saturation magnetization of about $1\mu_{\rm B}$ per Ni atom, indicating
that there are negligible magnetic dead layers at the interfaces. The $I_{\rm
c}R_{\rm n}(d_{\rm F})$ dependence of the SIFS junctions can be well described
by the dirty limit theory of Buzdin {\em et al.}~\cite{BuzdinPRB2003} yielding
the single characteristic length scale $\xi_{\rm F} = 3.88$\,nm for the decay
and oscillation of the critical current. From the measured IVCs and the Fiske
resonances appearing at finite applied magnetic flux the junction quality
factor has been determined in the wide frequency range between about 10 and
400\,GHz. At low frequencies the quality factor increases about linearly with
frequency due to the about frequency independent damping related to
quasiparticle tunneling, whereas it decreases proportional to $1/\omega$ at
high frequencies due to the increasing surface resistance of the junction
electrodes. The achieved quality factors range between about 5 at the junction
plasma frequency and 30 at about 200\,GHz and are sufficient for applications
of SIFS junctions in superconducting quantum circuits or experiments on
macroscopic quantum tunneling.

\begin{acknowledgement}
We gratefully acknowledge financial support by the Deutsche
Forschungsgemeinschaft via SFB~631 and the German Excellence Initiative via the
Nanosystems Initiative Munich (NIM). We thank M. Weides and J. Pfeiffer for
fruitful discussions as well as Th. Brenninger for technical support.
\end{acknowledgement}

\bibliographystyle{epj}
\bibliography{Wild_EPJB2010}
\end{document}